\documentclass[twoside,onecolumn]{IEEEtran}
\usepackage[latin10]{inputenc}
\usepackage{amsmath}
\usepackage{amsthm}
\usepackage{amssymb}
\usepackage{graphicx}
\usepackage{esint}

\usepackage{float}
\usepackage[caption = false]{subfig}

\makeatletter
\theoremstyle{plain}
\newtheorem{thm}{\protect\theoremname}

\providecommand{\theoremname}{Theorem}

\makeatother

\providecommand{\theoremname}{Theorem}

\begin{document}

\title{Regularized Estimation of Information via High Dimensional Canonical Correlation Analysis}

\author{Jaume~Riba,~\IEEEmembership{Senior Member,~IEEE} and Ferran~de Cabrera,~\IEEEmembership{Student Member,~IEEE} 
\thanks{
This work is supported by project TEC2016-76409-C2-1-R (WINTER), ``Ministerio de Economia y Competividad'', Spanish National Research Plan, by the Catalan administration (AGAUR) 			under 2017-SGR-578, and fellowship FI 2019 by the Secretary for University and Research of the ``Generalitat de Catalunya" and the European Social Fund.

Jaume~Riba and Ferran~de Cabrera are with the Signal Theory and Communications Department, Technical University of Catalonia (UPC), Barcelona 08034, Spain (e-mail: 					jaume.riba@upc.edu; ferran.de.cabrera@upc.edu).

This paper was presented in part at the 44th International Conference on Acoustics, Speech, and Signal Processing, ICASSP 2019.

This  work is a preprint version and it has been submitted to the IEEE for possible publication.

}}

\maketitle
\markboth{}{}
\begin{abstract}
In recent years, there has been an upswing of interest in estimating information from data emerging in a lot of areas beyond communications. This paper aims at estimating the information between two random phenomena by using consolidated second-order statistics tools. The squared-loss mutual information is chosen for that purpose as a natural surrogate of Shannon mutual information. The rationale for doing so is developed for i.i.d. discrete sources -mapping data onto the simplex space-, and for analog sources -mapping data onto the characteristic space-, highlighting the links with other well-known related concepts in the literature based on local approximations of information-theoretic measures. The proposed approach gains in interpretability and scalability for its use on large datasets, providing physical interpretation to the free regularization parameters. Moreover, the structure of the proposed mapping allows resorting to Szeg\"o's theorem to reduce the complexity for high dimensional mappings, exhibiting strong dualities with spectral analysis. The performance of the proposed estimators is analyzed using Gaussian mixtures.
\end{abstract}



\section{Introduction}

Entropy and mutual information, introduced by Shannon in 1948, are well-known concepts with clear operational significance in the field of information theory and communications: they establish fundamental limits in data compression and data transmission \cite{cover-2006}. More generally, Kullback-Leibler (KL) divergence (also called relative entropy) is a dissimilarity measure between distributions, being mutual information just a particular case. In the last decades, researchers have used the concepts of entropy, mutual information and divergence in a wide class of areas beyond communications, such as data science, machine learning, neuroscience, economics, biology, language and other experimental sciences. In these areas, the aforementioned concepts have proven their utility as tools for measuring randomness, dependence and similarity of random phenomena \cite{wang_2009,verdu-2019}, substituting or working together with the conventional statistical tools of variance and covariance. As a prominent example, the field of information-theoretic learning \cite{principe_2010} cuts across signal processing and machine learning by looking at machine learning under the umbrella of information theory. This new perspective for knowledge discovering provides guidelines for the design of nonparametric universal tools for data analytics \cite{kwang-2015}. Especially, the wish for interpretability and for understanding the learning process has become a challenging aspect in practical applications of machine learning systems due to their lack of ability to explain their actions to humans \cite{eldar-2017}. Although these models exhibit impressive capabilities, the development of tools for measuring information, which is the main motivation of this paper, can help them in reducing their vulnerability to attacks, and can provide means for diagnosis in the case of failures.


\subsection{Main contributions, related works and overall organization}

By delving into fundamental concepts of information theory and statistical signal processing, this paper aims at developing insightful tools for measuring meaningful indicators of the amount of information contained in raw data with two main objectives in mind: endowing the methods with as much interpretability as possible, thus providing insight on the selection of their free parameters; and leveraging as much as possible classical and consolidated statistical signal processing techniques based on second-order statistics.

The literature on empirical estimation of information measures and its applications is long, and a guide has been recently provided in \cite{verdu-2019}. The main contributions of this work are the following.
\begin{enumerate}
\item Providing a fresh view of the squared-loss mutual information surrogate for both discrete and analog sources. Its relation with other information measures is investigated and contextualized in a unified manner, focusing on the idea of local approximations. 
\item Linking the problem of estimation of information with the classical problem of Canonical Correlation Analysis (CCA), which is used in many fields of statistical signal processing. 
\item The proposal of an explicit universal mapping from analog sources to complex steering vectors, leading to a computationally efficient alternative to kernel-CCA (KCCA) methods and to their mechanism for regularization.
\item Providing interpretability and insight to the problem of estimating information and to its regularization.
\item The proposal of a reduced complexity approximate estimator resorting to the asymptotic behavior of Toeplitz matrices.
\end{enumerate}
Concerning the related works, the presented approach is an extension of the main ideas shortly provided by the authors in \cite{cabrera-2019}. Although the interest on surrogates of entropy, KL divergence and mutual information, such as R\'enyi entropy, R\'enyi divergence, $f$-divergence and chi-squared ($\chi^{2}$) divergence has a long and rich history (see \cite{verdu-2019} and references therein), its use for data analytics has been particularly focused in \cite{principe_2010}. The idea of local approximations of information measures exposed in this paper is very similar to the linear information coupling approach proposed in \cite{huang-2015,kwang-2015}, which was used there as a tool for developing insights on otherwise intractable problems in the field of communications. Compared with the kernel-CCA approach \cite{bach-2002} that uses the dual form of the models (kernel trick) -thus precluding its direct use in applications involving large datasets-, the proposed alternative stays in the primal model, thus gaining intuition and scalability of the overall data processing. Moreover, in \cite{Braun_relevant_dimensions_2008} it is suggested that the right choice of mapping function leads to a better representation of the data in the feature space, enabling to capture as much information as possible with a reduced dimensional mapping. The proposed statistics based on a Frobenius norm of a coherence matrix is very related to the local test proposed in \cite{ramirez-2013} for Gaussian vectors, with the difference that our result applies for any kind of data mapped on a specific feature space. The regularization idea based on Gaussian convolutions is inspired on \cite{goldfeld_2019}. Finally, the use of the Szeg\"o's theorem exploiting the analogy between a probability density function and a power spectral density was also explored in \cite{ramirez-2009} for Kullback-Leibler divergence estimation, by using autoregressive models for the densities.

This paper is organized as follows. Section II presents a unified overview of information theoretic surrogates, providing an original and fresh description of links among different approaches, and finishing with a short outline of the proposed overall strategy. Then, Section III focuses on discrete sources and shows the fundamental link between the proposed surrogate and classical second-order statistics. Once the structure of the problem is unveiled, Section IV moves to the extension to analog sources along with insightful tools for regularization and complexity reduction. The performance of the proposed estimators is illustrated by computer simulations in Section V and VI summarizes the main conclusion of this work. 


\subsection{Notation}

Column vectors: bold-faced lower case letters. Matrices: bold-faced upper case letters. $[\mathbf{\mathbf{A}}]_{n,m}$: element at the $n$-th row and $m$-th column of matrix $\mathbf{A}$. $[\mathbf{a}]_{n}$: $[\mathbf{\mathbf{a}}]_{n,1}$. $[\mathbf{a}]$: diagonal matrix with diagonal elements $[[\mathbf{a}]]_{n,n}=[\mathbf{a}]_{n}$. $(.)^{T}$: transpose. $(.)^{H}$: Hermitian transpose. $\text{tr}(\mathbf{A})$: trace. $||.||$: Frobenius or Euclidean norm of a vector, matrix or function. $|.|$: absolute value of a complex number, or cardinality of a set. $\mathbf{A}\in\mathbb{R}^{N\times M}$: real matrix of dimension $N\times M$. $\mathbf{A}\in\mathbb{C}^{N\times M}$: complex matrix of dimension $N\times M$. $\mathbf{a}\in\mathbb{R}^{N}$: $\mathbf{a}\in\mathbb{R}^{N\times1}$. $\mathbf{a}\in\mathbb{C}^{N}$: $\mathbf{a}\in\mathbb{C}^{N\times1}$. $\mathbb{R}_{+}$: set of positive real numbers. $\mathbf{x}\sim\mathcal{N}(\boldsymbol{\mu},\mathbf{R})$: $\mathbf{x}$ is a real Gaussian random vector of mean $\boldsymbol{\mu}$ and covariance matrix $\mathbf{R}$. $\mathbf{x}\sim\mathcal{\mathcal{CN}}(\boldsymbol{\mu},\mathbf{R})$: $\mathbf{x}$ is a complex Gaussian random vector of mean $\boldsymbol{\mu}$ and covariance matrix $\mathbf{R}$. $\mathbb{E}_{p}$: statistical expectation operator ($\mathbb{E}_{p}[f(x)]=\int fdP$), where $p$ is the mass function (or density function for analog variables), and $P$ is the probability measure (or the cumulative distribution function for analog variables). $\left\langle x(l)\right\rangle _{L}=L^{-1}\sum_{l=1}^{L}x(l)$: $L$-th length sample mean operator. $\mathbf{I}_{D}$: $D\times D$ identity matrix. $\mathbf{0}_{D}$: $D\times1$ all-zeros vector. $\mathbf{1}_{D}$: $D\times1$ all-ones vector. $1_{a}$: indicator function ($1_{a}=1$ if $a$ is true, and $1_{a}=0$, otherwise). $\mathbf{A}^{1/2}$: Hermitian square root matrix of the Hermitian matrix $\mathbf{A}$. $\mathbf{A}^{-1/2}$: Hermitian square root matrix of the Hermitian matrix $\mathbf{A}^{-1}$. $\mathbf{a}^{\alpha}$: element-wise power of a vector. $\mathbf{a}^{T\alpha}=\left(\mathbf{a}^{\alpha}\right)^{T}$. $\text{Toe}(\mathbf{c})$: Toeplitz-Hermitian matrix constructed from its first column $\mathbf{c}$. $\odot$: Hadamard product. $*$: convolution operator. $\delta_{mn}$: Kronecker delta. $\left\lceil x\right\rceil $: ceiling function.


\section{Information-theoretic measures for data analytics}

Shannon entropy and Kullback-Leibler (KL) divergence are fundamental quantities in information theory and its applications \cite{cover-2006}. Some surrogates of these quantities have been proposed for data analytics with the goal of simplifying the estimation process. In this section we provide a unified rationale for the derivation of surrogate information measures (entropy, divergence and mutual information) in a way as natural as possible, along with more details concerned with the related work, and we finish with a summary of the key ideas and more concrete goals of this paper. Through all the paper, we will assume that $p_{X}(x)$, defined on a set $\mathcal{X}$, is either a mass function (for discrete sources) or a square-integrable density function (for continuous sources) associated to the random variable $X$.


\subsection{Information potential surrogate }

The Shannon entropy (in \textit{nats}) is defined as
\begin{equation}
H(p_{X})=-\mathbb{E}_{p_{X}}\ln p_{X}(x).
\end{equation}
A natural surrogate of Shannon entropy can be obtained by applying the Jensen inequality. In particular, as $\ln(.)$ is concave, we obtain 
\begin{equation}
H(p_{X})=-\mathbb{E}_{p}\ln p_{X}(x)\geq-\ln\mathbb{E}_{p_{X}}p_{X}(x)=H_{2}(p_{X}),\label{h2}
\end{equation}
with equality if and only if the source is uniform; otherwise, we get an strict inequality ($>$) as a consequence of the strict concavity of $\ln(.)$. The right-hand term $H_{2}(p_{X})$ in (\ref{h2}) is just the second-order R\'enyi entropy, whose use for estimation via kernel methods has been explored in \cite{principe_2010}. Second-order R\'enyi entropy can also be expressed as follows:
\begin{equation}
H_{2}(p_{X})=-\ln\left(1-S_{2}\left(p_{X}\right)\right),
\end{equation}
where $S_{2}\left(p_{X}\right)$ is the Tsallis entropy of second-order entropic index given by
\begin{equation}
S_{2}\left(p_{X}\right)=1-V_{2}(p_{X}),
\end{equation}
being $V_{2}(p_{X})$ defined as the \textit{information potential}:
\begin{equation}
V_{2}(p_{X})=\mathbb{E}_{p_{X}}p_{X}(x)=\left\Vert p_{X}\right\Vert ^{2}.\label{IPd}
\end{equation}
From the fundamental logarithm inequality $\ln(1+x)\leq x$ we can state that the second-order Tsallis entropy lower bounds the second-order R\'enyi entropy, that is
\begin{equation}
H(p_{X})\geq H_{2}(p_{X})\geq S_{2}\left(p_{X}\right).
\end{equation}
The information potential $V_{2}(p_{X})$, as defined in (\ref{IPd}), is just the squared Euclidean norm of the mass function or density function of the source, and it admits physical interpretations \cite{principe_2010}. In the discrete case, the information potential is also called the collision probability \cite{fehr-2014} as it represents the probability that two independent outcomes of the source are equal. In the general case, the information potential has been used as a natural surrogate of (reversed sign) entropy because no logarithm is involved in its definition \cite{principe_2010}. The information potential admits an estimation procedure from data which is much more natural than trying to estimate entropy itself. In particular, if one formulates a plug-in estimation method by first estimating the density function via the Parzen\textendash Rosenblatt window method \cite{parzen-1962,rosenblatt-1956}, a final estimator is obtained by resorting to kernel methods \cite{principe_2010,seth-2011}. The information potential has also been proposed in \cite{cabrera2017} and \cite{cabre-2017} as a means to obtain robustness to outliers in the estimation of determinants of covariance matrices. For a review of plug-in methods and other methods for entropy estimation, the reader is referred to   \cite{wang_2009}.
 

\subsection{Chi-squared divergence surrogate }

Inspired by the above rationale for the derivation of a natural surrogate of entropy, we next proceed with the divergence concept in a similar way. The KL divergence (in \textit{nats}) between two probability mass or density functions $p_{X}(x)$ and $q_{X}(x)$, defined on the same set $\mathcal{X}$, is given by
\begin{equation}
D\left(p_{X}||q_{X}\right)=\mathbb{E}_{p_{X}}\ln\frac{p_{X}(x)}{q_{X}(x)}.\label{kldef}
\end{equation}
For continuous random variables, absolute continuity of the densities with respect to each other is assumed, leading to bounded KL divergence. KL divergence is non-negative, and it is zero if and only if $p_{X}(x)=q_{X}(x)$. A natural surrogate of KL divergence can be obtained by applying the Jensen inequality to (\ref{kldef}). In particular, as $\ln(.)$ is concave, we obtain
\begin{equation}
D\left(p_{X}||q_{X}\right)\leq\ln\mathbb{E}_{p_{X}}\frac{p_{X}(x)}{q_{X}(x)}=D_{2}\left(p_{X}||q_{X}\right),
\end{equation}
with equality if and only if $p_{X}(x)=q_{X}(x)$. For $p_{X}(x)\neq q_{X}(x)$, i.e. for non-zero KL divergence, we get an strict inequality ($<$) as a consequence of the strict concavity of $\ln(.)$. The right-hand term $D_{2}\left(p_{X}||q_{X}\right)$ is just the second-order R\'enyi divergence for which some operational characterization have been provided \cite{erven-2014}.  R\'enyi divergence belongs to the class of $f$-divergences, which are useful, for example, in pattern recognition applications to identify independent components \cite{bao-2003}, as dissimilarity
measures for image registration \cite{pluim-2004}, and for target tracking \cite{Kreucher-2003}. Second-order R\'enyi divergence can also be expressed as follows:
\begin{equation}
D_{2}\left(p_{X}||q_{X}\right)=\ln\left(1+D_{\chi^{2}}\left(p_{X}||q_{X}\right)\right),\label{explicit}
\end{equation}
where $D_{\chi^{2}}\left(p_{X}||q_{X}\right)$ is the Pearson chi-squared divergence given by 
\begin{equation}
D_{\chi^{2}}\left(p_{X}||q_{X}\right)=\mathbb{E}_{p_{X}}\frac{p_{X}(x)}{q_{X}(x)}-1 =\mathbb{E}_{p_{X}}\left(\frac{p_{X}(x)-q_{X}(x)}{\sqrt{p_{X}(x)q_{X}(x)}}\right)^{2}=\left\Vert \frac{p_{X}-q_{X}}{\sqrt{q_{X}}}\right\Vert ^{2},\label{chisq_disc}
\end{equation}
(see Appendix A). From the fundamental logarithm inequality $\ln(1+x)\leq x$ we can state that the Pearson chi-squared divergence upper bounds the second-order R\'enyi divergence, that is
\begin{equation}
D\left(p_{X}||q_{X}\right)\leq D_{2}\left(p_{X}||q_{X}\right)\leq D_{\chi^{2}}\left(p_{X}||q_{X}\right),\label{ineqs-div}
\end{equation}
where the inequalities are strict ($<)$ for non-zero divergence, and they become equality (to zero) if and only if $p_{X}(x)=q_{X}(x)$. Looking at (\ref{ineqs-div}), it is worth noting that both R\'enyi and chi-squared divergences may be infinite even for finite KL divergence. As an example, we can mention the case of Gaussian $p_{X}$ and $q_{X}$ distributions with the variance of $p_{X}$ being more than twice the variance of $q_{X}$ (see \cite{erven-2014}, Eq. (10)). This issue is ignored in this paper because, as the focus is data analytics, the challenging problem is that of measuring divergence when it is small, as explained later on under the view of local approximations. Moreover, the empirical estimators will need to be ultimately regularized to cope with the limited data size, as detailed later on, which will lead to finite estimates for all scenarios.

As $D_{2}$ is an explicit and monotonic function of $D_{\chi^{2}}$ given in (\ref{explicit}), there is no practical difference between them in terms of computational complexity from data. For this reason, especially for clarity, we will focus only on the chi-squared divergence along this paper, having in mind that a tighter upper bound $D_{2}$ can be obtained from $D_{\chi^{2}}$ via (\ref{explicit}) if it were required for a particular application. Computing $D_{2}$ from $D_{\chi^{2}}$ may also be interesting in order to recover the additivity property of the obtained divergence measure with respect to independent (i.e. multiplicative) components in either $p_{X}(x)$ or $q_{X}(x)$, because while KL and R\'enyi divergence satisfy this property, the chi-squared divergence does not. A final consequence of the one-to-one relationship between R\'enyi and the adopted chi-squared divergence is that the chi-squared divergence inherits the invariance property of the R\'enyi divergence to nonlinear invertible transformations of the data. This follows from the more general data processing inequality (see, e.g., \cite{liese-2006}).

We propose the use of the chi-squared divergence as defined in (\ref{chisq_disc}) as a natural upper bound to the KL divergence that exhibits significant computational advantages for data analytics. Its main advantage comes from the fact that no logarithm is involved on the quantity $p_{X}dP_{X}/q_{X}$ that forms the integrand in the proof of (\ref{chisq_disc}) (see Appendix A). In any case, the logarithm is located outside the sum (or integral) if R\'enyi divergence is computed. This fact is what allows the use of second-order analysis techniques that are well known in statistical signal processing. The price to give entrance to these techniques is the need of mapping the data onto a high dimensional space, which constitutes the core idea explored in forthcoming sections. 


\subsubsection{Local approximation of KL divergence}

The selection of the chi-squared divergence as a natural surrogate of KL divergence can be further reasoned by means of the following alternative rationale. Consider that $p_{X}(x)$ and $q_{X}(x)$ are close to each other, that is $q_{X}(x)=p_{X}(x)+\epsilon\Delta(x)$ for some small quantity $\epsilon$, where $\Delta(x)$ is defined on the set $\mathcal{X}$ and constrained to have null area. Using the Taylor expansion of $\ln((1+\alpha)^{-1})$ up to the second order, i.e. $-\alpha+\alpha^{2}/2+O(\alpha^{3})$, we can write the KL divergence in (\ref{kldef}) as
\begin{equation}
D\left(p_{X}||p_{X}+\epsilon\Delta\right)=\mathbb{E}_{p_{X}}\ln\frac{p_{X}(x)}{p_{X}(x)+\epsilon\Delta(x)}=-\epsilon\mathbb{E}_{p_{X}}\left[\frac{\Delta(x)}{p_{X}(x)}\right]+\frac{1}{2}\epsilon^{2}\mathbb{E}_{p_{X}}\left[\left(\frac{\Delta(x)}{p_{X}(x)}\right)^{2}\right]+O(\epsilon^{3}).
\end{equation}
The first term is null since $\Delta(x)$ sums up zero, which implies that 
\begin{equation}
D\left(p_{X}||p_{X}+\epsilon\Delta\right)=\frac{1}{2}\epsilon^{2}\mathbb{E}_{p_{X}}\left[\left(\frac{\Delta(x)}{p_{X}(x)}\right)^{2}\right]+O(\epsilon^{3}).\label{aprox1}
\end{equation}
Let us now examine the local behavior of the chi-squared divergence. Using the Taylor expansion of $(1+\alpha)^{-1}$ up to the first order, i.e. $1-\alpha+O(\alpha^{2})$, we can write the chi-squared divergence in (\ref{chisq_disc}) as 
\begin{equation*}
D_{\chi^{2}}\left(p_{X}||p_{X}+\epsilon\Delta\right)=\mathbb{E}_{p_{X}}\frac{\left(-\epsilon\Delta(x)\right)^{2}}{p_{X}(x)\left(p_{X}(x)+\epsilon\Delta(x)\right)}=\epsilon^{2}\mathbb{E}_{p_{X}}\left[\left(\frac{\Delta(x)}{p_{X}(x)}\right)^{2}\left(1-\frac{\epsilon\Delta(x)}{p_{X}(x)}+O(\epsilon^{2})\right)\right]
\end{equation*}
\begin{equation}
=\epsilon^{2}\mathbb{E}_{p_{X}}\left[\left(\frac{\Delta(x)}{p_{X}(x)}\right)^{2}\right]+O(\epsilon^{3}).\label{aprox2}
\end{equation}
From (\ref{aprox1})\&(\ref{aprox2}), the following fundamental result can be stated: 
\begin{equation}
D\left(p_{X}||p_{X}+\epsilon\Delta\right)=\frac{1}{2}D_{\chi^{2}}\left(p_{X}||p_{X}+\epsilon\Delta\right)+O(\epsilon),\label{local}
\end{equation}
which means that half of the chi-squared divergence, that is $\frac{1}{2}D_{\chi^{2}}$, constitutes a local approximation of KL divergence for close distributions. This observation is important because while $D_{\chi^{2}}$ upper bounds KL divergence, $\frac{1}{2}D_{\chi^{2}}$ is instead a local approximation, but not an upper bound. Note finally that, as pointed out in \cite{huang-2015}, $D\left(p_{X}||p_{X}+\epsilon\Delta\right)$ and $D\left(p_{X}+\epsilon\Delta||p_{X}\right)$ are considered to be equal up to first order approximation, and the same happens for $D_{\chi^{2}}\left(p_{X}||p_{X}+\epsilon\Delta\right)$ and $D_{\chi^{2}}\left(p_{X}+\epsilon\Delta||p_{X}\right)$. 

As a related work, it is worth mentioning that local approximations of the KL divergence have been explored in \cite{huang-2015} under the context of Linear Information Coupling (LIC) problems and Euclidean information theory, motivated by the goal of translating information theory problems into linear algebra problems, thus avoiding computational and mathematical bottlenecks. Similarly, the present paper extends this crucial idea to the analog case with the goal of translating the problem of measuring statistical dependence into much more manageable second-order analysis problems. 


\subsection{Squared-loss mutual information surrogate}

Mutual information (MI) is an important concept in information theory that quantifies the statistical dependence between two random sources $X$ and $Y$, possibly defined on different sets $\mathcal{X}$ and $\mathcal{Y}$, respectively. The Shannon MI is defined as the KL divergence between the joint distribution and the product of the marginal distributions, both defined on the product set $\mathcal{X}\times\mathcal{Y}$:
\begin{equation}
I(X;Y)=D\left(p_{XY}||p_{X}p_{Y}\right)=\mathbb{E}_{p_{XY}}\ln\frac{p_{XY}(x,y)}{p_{X}(x)p_{Y}(y)}.
\end{equation}
Following the Jensen inequality upper bound idea examined in the previous subsection, a natural surrogate of mutual information for data analytics can be defined as follows:
\begin{equation}
I_{2}(X;Y)=D_{2}\left(p_{XY}||p_{X}p_{Y}\right)=\ln\mathbb{E}_{p_{XY}}\frac{p_{XY}(x,y)}{p_{X}(x)p_{Y}(y)}.\label{define_I2}
\end{equation}

The definition obtained in (\ref{define_I2}) is in agreement with the definition of R\'enyi mutual information proposed in \cite{tridenski-2015}. However, it has to be noted that there are other possible ways to accomplish the generalization from the R\'enyi divergence to R\'enyi mutual information, most notably those suggested by Arimoto, Csisz\'ar and Sibson (see \cite{verdu-2015} for a short review). Other definitions, as those in \cite{Tomamichel-2018} and \cite{lapidoth-2019}, involve a minimization process with respect to the marginals, motivated by operational interpretations. 

As a particular case of (\ref{explicit})\&(\ref{chisq_disc}), the second-order mutual information defined in (\ref{define_I2}) can be written as
\begin{equation}
I_{2}(X;Y)=\ln\left(1+I_{s}\left(X;Y\right)\right),
\end{equation}
where
\begin{equation}
I_{s}\left(X;Y\right)=\mathbb{E}_{p_{XY}}\frac{p_{XY}(x,y)}{p_{X}(x)p_{Y}(y)}-1=\mathbb{E}_{p_{XY}}\left(\frac{p_{XY}(x,y)-p_{X}(x)p_{Y}(y)}{\sqrt{p_{XY}(x,y)p_{X}(x)p_{Y}(y)}}\right)^{2}=\left\Vert \frac{p_{XY}-p_{X}p_{Y}}{\sqrt{p_{X}p_{Y}}}\right\Vert ^{2} \label{smi}
\end{equation}
is the squared-loss mutual information (SMI) introduced in \cite{suzuki-2009} for feature selection, and it is just the Pearson chi-squared divergence from $p_{XY}(x,y)$ to $p_{X}(x)p_{Y}(y)$. It is also worth mentioning that the SMI can also be deduced from the called \textit{normalized cross-covariance operator} (see \cite{Fukumizu_conditional}, Eq. (9)). In this case, while constructing the Hilbert-Schmidt norm of a covariance operator through kernel methods, the associated explicit kernel-free integral expression corresponds to the SMI. This alternative way is particularly interesting since establishes a clear link between second-order statistics and the SMI, provided that the data is mapped onto certain feature spaces, as we will see in the next section.
Finally, note from (\ref{ineqs-div}) that 
\begin{equation}
I(X;Y)\leq I_{2}(X;Y)\leq I_{s}(X;Y),
\end{equation}
where the inequalities are strict ($<)$ for dependent sources, and they become equality (to zero) if and only if the sources are independent. 


\subsubsection{Local approximation of Shannon mutual information}

Using the local approximation of the KL divergence described in (\ref{local}), if we assume the case of low dependence, that is $p_{XY}(x,y)=p_{X}(x)p_{Y}(y)+\epsilon\Delta(x,y)$ for some small quantity $\epsilon$, where $\Delta(x,y)$ is defined on the set $\mathcal{X}\times\mathcal{Y}$ and constrained to have null area, we can state that
\begin{equation}
I\left(X;Y\right)=\frac{1}{2}I_{s}(X;Y)+O(\epsilon), \label{local-1}
\end{equation}
as a particular case of (\ref{local}), which means that half of the squared-loss mutual information, that is $\frac{1}{2}I_{s}$, constitutes a local approximation of Shannon mutual information for low dependence scenarios. Once again, this observation is important because while $I_{s}$ upper bounds mutual information, $\frac{1}{2}I_{s}$ is instead a local approximation, but not an upper bound. As a simple example, for the case that $p_{XY}(x,y)$ is a bivariate normal density with Pearson coefficient $\rho=\text{cov}(X,Y)/(\sigma_{X}\sigma_{Y})$, we have $I\left(X;Y\right)=-0.5\ln(1-\rho^{2})$ and $0.5I_{s}\left(X;Y\right)=0.5\rho^{2}/(1-\rho^{2})$, which are equal up to first order approximation. 


\subsubsection{Other quadratic dependence measures}

Among other possibilities, a non-negative dependence measure that satisfies the requirement of being zero if and only if the sources are independent can also be defined as follows (see \cite{seth-2011}, Eq. (4)): 
\begin{equation*}
\xi(X;Y)=\mathbb{E}_{p_{XY}}\left(\frac{p_{XY}(x,y)-p_{X}(x)p_{Y}(y)}{\sqrt{p_{X,Y}(x,y)}}\right)^{2}
\end{equation*}
\begin{equation}
=\left\Vert p_{XY}-p_{X}p_{Y}\right\Vert ^{2}. \label{smi-1-1}
\end{equation}
Although this measure looks simpler, as it becomes the squared norm of a difference of densities, it lacks connection with Shannon mutual information, since nor inequalities nor local behavior can be stated in the way that have been established in (\ref{local-1}) for the squared-loss mutual information measure $I_{s}\left(X;Y\right)$ defined in (\ref{smi}). 


\subsection{Summary of the key idea}

\begin{figure}[t]
\centering
  \includegraphics[clip,scale=0.7]{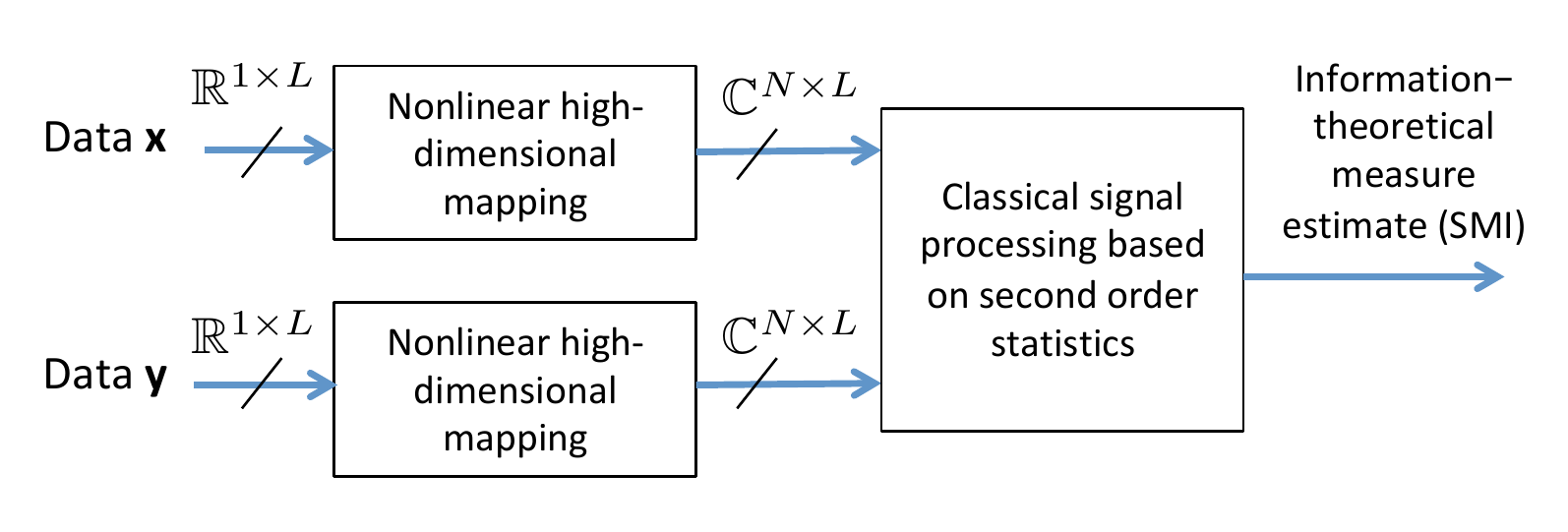}
\caption{Block diagram of the proposed data analytics strategy. \label{fig1}}
 \end{figure}

After the short review and unified rationale presented above, we can summarize the goal of this paper. We have derived the surrogates (\ref{chisq_disc}) and (\ref{smi}) for divergence and mutual information, respectively, which share the following important properties.
\begin{enumerate}
\item The surrogates are upper bounds of information theoretic measures of well-known operational meaning, namely Kullback-Leibler divergence and Shannon mutual information. By being upper bounds, we make sure that relevant information will not be lost by using the surrogates for data analytics.
\item Half of the magnitudes measured by the surrogates are local approximations of meaningful information measures, which implies that they adopt meaningful values for the critical scenarios of close distributions for the divergence case, and small dependence regime for the mutual information case. Thus, the good local approximation property of the surrogates ensures that, in the challenging and interesting cases of measuring small information, the magnitude measured has full meaning. This behavior is lost by other quadratic measures of information. 
\item The surrogates can be expressed as a second-order moment, that is, as the expectation of the squared of a random variable involving solely a ratio of densities without any logarithm. The implication is that, by designing adequate pre-conditioning of the data, classical second-order analysis techniques should be enough for estimating information.
\end{enumerate}
The purpose of what follows is to propose a universal mapping strategy from the data onto a high dimensional feature space, such that the information can be extracted from that space by standard second-order signal processing techniques. The ultimate goal is to provide a rationale for the two-step data analytics strategy depicted in Fig. \ref{fig1}. Basically, the purpose of the first stage is to analyze complex dependencies between two data sources by first mapping $L$ samples of the bivariate data onto a high-dimensional space defined on the complex field. The dimension $N$ should be high enough to make sure that the maximum amount of complex associations potentially present on the data are captured, but it should be sufficiently small to provide reasonable computational complexity as well as regularization capabilities. After that, the second stage is based on second-order analysis techniques focused on describing linear dependencies between sets of variables. For instance, CCA ensures the previous statement and is a well known technique in the statistical signal processing field.

\section{Discrete sources: second-order statistics on the simplex feature space}

We first focus our attention on discrete sources since the key bridge to relate information with second-order statistics emerges more clearly in this case. Later, we will leverage this idea in order to smoothly generalize the concept to the more challenging case of analog sources.

Consider that $X$ and $Y$ are discrete random variables with alphabets $\mathcal{X}=\{x_{n}\}_{n=1,2,\ldots,N}$ and $\mathcal{Y}=\{y_{m}\}_{m=1,2,\ldots,M}$, respectively. Let us define the marginal probability column vectors $\tilde{\boldsymbol{p}}\in\mathbb{R}_{+}^{N}$ and $\tilde{\boldsymbol{q}}\in\mathbb{R}_{+}^{M}$ as $[\tilde{\boldsymbol{p}}]_{n}=\Pr\{X=x_{n}\}=p_{X}(x_{n})$ for $n=1,2,\ldots,N$ and $[\tilde{\boldsymbol{q}}]_{m}=\Pr\{Y=y_{m}\}=p_{Y}(y_{m})$ for $m=1,2,\ldots,M$. Similarly, we define the joint probability matrix $\tilde{\mathbf{J}}\in\mathbb{R}_{+}^{N\times M}$ as $[\tilde{\mathbf{J}}]_{n,m}=\Pr\{X=x_{n};Y=y_{m}\}=p_{XY}(x_{n},y_{m})$. Then, the SMI defined in (\ref{smi}) can be expressed as follows:
\begin{equation}
I_{s}\left(X;Y\right)=\sum_{n=1}^{N}\sum_{m=1}^{M}[\tilde{\mathbf{C}}]_{n,m}^{2}=\text{tr}(\tilde{\mathbf{C}}^{T}\tilde{\mathbf{C}})=||\mathbf{\tilde{C}}||^{2},\label{smitrace}
\end{equation}
where 
\begin{equation}
\tilde{\mathbf{C}}=[\tilde{\mathbf{p}}]^{-1/2}(\tilde{\mathbf{J}}-\tilde{\mathbf{p}}\tilde{\mathbf{q}}^{T})[\tilde{\mathbf{q}}]^{-1/2}.\label{cohe}
\end{equation}
Matrix $\tilde{\mathbf{C}}\in\mathbb{R}_{+}^{N\times M}$ in (\ref{cohe}) will be referred to as \textit{coherence} matrix for the reasons explained later on, particularly due to its intimate link with the well known CCA tool in statistical signal processing. Moreover, the form of this matrix is encountered as well in the areas of information theory under the context of Linear Information Coupling (LIC) problems and Hirschfeld-Gebelein-R\'enyi (HGR) maximal correlation concept. As (\ref{cohe}) contains the key ideas explored in this paper, we next provide an overview of these links along with new statements from known notions, all concerned with the problem of estimating information.


\subsection{Relation to Linear Information Coupling (LIC) problems}

Matrix $\tilde{\mathbf{C}}$ in (\ref{cohe}) can be expressed as follows:
\begin{equation}
\tilde{\mathbf{C}}=\left(\mathbf{B}-\tilde{\mathbf{q}}^{1/2}\tilde{\mathbf{p}}{}^{T/2}\right)^{T},\label{cebe}
\end{equation}
where 
\begin{equation}
\mathbf{B}=[\tilde{\mathbf{q}}]^{-1/2}\tilde{\mathbf{J}}^{T}[\tilde{\mathbf{p}}]^{-1/2}.\label{labe}
\end{equation}
To provide an interpretation of (\ref{labe}), consider that the source $Y$ is the output of a discrete memory-less channel whose input is $X$. Let $\mathbf{W}\in\mathbb{R}_{+}^{M\times N}$ be the channel transition matrix defined by the conditional probabilities of the outputs given the inputs, that is, 
\begin{equation}
[\mathbf{W}]_{m,n}=\Pr(Y=y_{m}|X=x_{n}).
\end{equation}
We can then write the elements of the joint mass function as $[\tilde{\mathbf{J}}]_{n,m}=\Pr(Y=y_{m}|X=x_{n})\Pr(X=x_{n})$ or, more compactly, 
\begin{equation}
\tilde{\mathbf{J}}^{T}=\mathbf{W}[\tilde{\mathbf{p}}].\label{jota}
\end{equation}
Using (\ref{jota}) in (\ref{labe}) we can write
\begin{equation}
\mathbf{B}=[\tilde{\mathbf{q}}]^{-1/2}\mathbf{W}[\tilde{\mathbf{p}}]^{1/2}.\label{DTM}
\end{equation}
This matrix is called the \textit{divergence transition matrix} (DTM) of a discrete channel \cite{huang-2012,huang-2015}, and it plays a fundamental role as a tool for translating information theory problems into linear algebra problems. Similarly with the approach followed in this paper, the linear algebra in LIC problems arises as well as result of a local approximation of the KL divergence, and provides rich insights, guidelines and geometrical interpretations in classical optimization problems encountered in the field of communications. In particular, it can be easily shown that the maximum singular value of the DTM is $\sigma_{1}=\sigma_{\text{max}}(\mathbf{B})=1$, corresponding to right and left singular vectors $\tilde{\mathbf{p}}{}^{1/2}$ and $\tilde{\mathbf{q}}{}^{1/2}$, respectively, \cite{huang-2015}. In fact, it is its \textit{second largest} singular value ($\sigma_{2}=\sigma_{\text{smax}}(\mathbf{B})$), along with the corresponding right and left singular vectors ($\mathbf{v}_{\text{s}}(\mathbf{B})$ and $\mathbf{u}_{\text{s}}(\mathbf{B})$), those that become useful and insightful for the optimization problems explored using the LIC approach. As a connection, the following theorem establishes the physical meaning of $\sigma_{\text{smax}}(\mathbf{B})$, $\mathbf{v}_{\text{s}}(\mathbf{B})$ and $\mathbf{u}_{\text{s}}(\mathbf{B})$ within the framework of this paper, namely the measure of statistical dependence:
\begin{thm}
\label{theo seclargest}Let $\{\lambda_{i}\}_{i=1:\min(N,M)}$ be the singular values of the coherence matrix $\tilde{\mathbf{C}}$ in (\ref{cohe}). Then: i) the minimum singular value is zero; ii) the largest singular value is equal to the second largest singular value of the divergence transition matrix in (\ref{DTM}); iii) the squared-loss mutual information in (\ref{smitrace}) is upper bounded by $\min(N,M)-1$.
\end{thm}
\begin{IEEEproof}
According to the mentioned properties of matrix $\mathbf{B}$, we can write its SVD as follows:
\begin{equation}
\mathbf{B}=\tilde{\mathbf{q}}^{1/2}\tilde{\mathbf{p}}{}^{T/2}+\sigma_{\text{smax}}\mathbf{u}_{\text{smax}}\mathbf{v}_{\text{smax}}^{T}+\sum_{i=3}^{\min(N,M)}\sigma_{i}\mathbf{u}_{i}\mathbf{v}_{i}^{T}.
\end{equation}
\vspace{8pt}Now, from the expression in (\ref{cebe}), it is clear that the SVD of matrix $\tilde{\mathbf{C}}$ is:
\begin{equation}
\tilde{\mathbf{C}}=\sigma_{\text{smax}}\mathbf{u}_{\text{smax}}\mathbf{v}_{\text{smax}}^{T}+\sum_{i=2}^{\min(N,M)-1}\lambda_{i}\mathbf{u}_{i+1}\mathbf{v}_{i+1}^{T},
\end{equation}
where $\sigma_{\text{smax}}(\mathbf{B})=\lambda_{\text{max}}(\tilde{\mathbf{C}})$. In short, we find that the second largest singular value of the divergence transition matrix (a fundamental quantity in LIC problems) is equal to the largest singular value of the coherence matrix (a fundamental quantity in measuring statistical dependence). Finally, as the eigenvalues of matrix $\mathbf{\tilde{C}}^{T}\mathbf{\tilde{C}}$ are the squared modulus of the singular values of $\mathbf{\tilde{C}}$, which are all smaller than $1$ and the minimum is $0$, we obtain the stated upper bound on the SMI.
\end{IEEEproof}


\subsection{Relation to Canonical Correlation Analysis (CCA)}

The above matrix $\tilde{\mathbf{C}}$ in (\ref{cohe}) has the form of a coherence matrix and, therefore, it turns out that the squared-loss mutual information $I_{s}\left(X;Y\right)$ can be directly related with the standard CCA method \cite{hotel-1936}. This connection of ideas is relevant since CCA is an important tool applied in many fields of signal processing and machine learning, so it should not be surprising if, eventually, that notion becomes fundamental as well for the problem of estimating information. Interestingly, the connection with CCA that will be unveiled in the sequel is a direct consequence of the fact that no logarithm is present in the definition of the squared-loss mutual information surrogate proposed in this paper. 

\begin{figure}[tp]
\centering
\includegraphics[clip,scale=0.6]{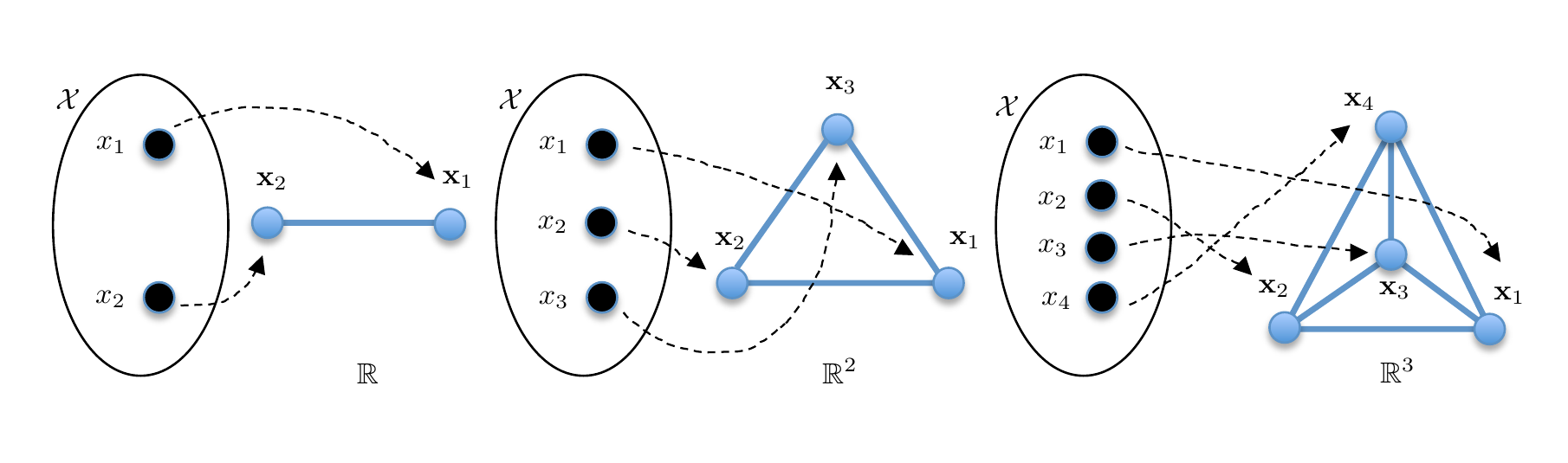}\caption{Illustration of the mapping $\mathcal{X}\rightarrow\mathbb{R}^{|\mathcal{X}|-1}$ onto the $(|\mathcal{X}|-1)$-\textit{simplex}. \label{fig2}}
\end{figure}

To see that bridge, let us try to express matrix $\tilde{\mathbf{C}}$ as a function of second-order statistics computed from the available data consisting of a sequence of $L$ i.i.d. pairs $\{x(l),y(l)\}\in\mathcal{X}\times\mathcal{Y}$ for $l=1,2,\ldots L$. Let $\hat{\tilde{\mathbf{p}}}$, $\hat{\tilde{\mathbf{q}}}$ and $\hat{\tilde{\mathbf{J}}}$ be estimates of the marginal and joint mass functions. From (\ref{smitrace})\&(\ref{cohe}), we define a plug-in estimator of the SMI as $\hat{I}_{s}\left(X;Y\right)=||\hat{\tilde{\mathbf{C}}}||_{F}^{2}$, where $\hat{\tilde{\mathbf{C}}}=[\hat{\tilde{\mathbf{p}}}]^{-1/2}(\hat{\tilde{\mathbf{J}}}-\hat{\tilde{\mathbf{p}}}\hat{\tilde{\mathbf{q}}}^{T})[\hat{\tilde{\mathbf{q}}}]^{-1/2}$. Let us define the full-rank \footnote{The data matrices are assumed full-rank for clarity, implying that $L$ is sufficiently large such that $(x_{n},y_{m})\in\{x(l),y(l)\}_{l=1:L}$ for all $n=1:N$ and $=1:M$. Note that $[\hat{\tilde{\mathbf{p}}}]$ and $[\hat{\tilde{\mathbf{q}}}]$ are therefore invertible under this assumption. The issue of rank-deficient data matrices will be specifically addressed later on.} data matrices $\mathbf{D}_{x}$ ($N\times L$) and $\mathbf{D}_{y}$ ($M\times L$) as follows:
\begin{equation}
[\mathbf{D}_{x}]_{n,l}=1_{x(l)=x_{n}},\qquad[\mathbf{D}_{y}]_{n,l}=1_{y(l)=y_{m}}.
\end{equation}
These data matrices are the result of a one-to-one mapping process from the elements of the sources to the canonical basis of dimension equal to the set cardinality. Clearly, the mass function estimates required by the plug-in estimator of SMI can be computed through first and second-order statistics as follows:
\begin{equation*}
\hat{\tilde{\mathbf{p}}}=\frac{1}{L}\mathbf{D}_{x}\boldsymbol{1},\qquad\hat{\tilde{\mathbf{q}}}=\frac{1}{L}\mathbf{D}_{y}\boldsymbol{1},
\end{equation*}
\begin{equation*}
[\hat{\tilde{\mathbf{p}}}]=\frac{1}{L}\mathbf{D}_{x}\mathbf{D}_{x}^{H},\qquad[\hat{\tilde{\mathbf{q}}}]=\frac{1}{L}\mathbf{D}_{y}\mathbf{D}_{y}^{H},
\end{equation*}
\begin{equation}
\hat{\tilde{\mathbf{J}}}-\hat{\tilde{\mathbf{p}}}\hat{\tilde{\mathbf{q}}}^{T}=\frac{1}{L}\mathbf{D}_{x}\mathbf{P}_{\mathbf{1}}^{\bot}\mathbf{D}_{y}^{H},\label{unas}
\end{equation}
where $\mathbf{P}_{\mathbf{1}}^{\bot}=\mathbf{I}-\boldsymbol{1}\boldsymbol{\mathbf{1}}^{T}/L$ is the projection matrix onto the orthogonal space spanned by $\boldsymbol{\mathbf{1}}$. As a result, the mass function estimates required in the computation of the SMI are just the two sample mean vectors, the two autocorrelation matrices and the cross-covariance matrix. The following theorem introduces a preliminary link with CCA:
\begin{thm}
\label{theo coherence}Preliminary link SMI-CCA: Let $\mathbf{X}\in\mathbb{C}^{N\times L}$ and $\mathbf{Y}\in\mathbb{C}^{M\times L}$ be data matrices obtained as $\mathbf{X}=\mathbf{F}\mathbf{D}_{x}$ and $\mathbf{Y}=\mathbf{G}\mathbf{D}_{y}$, respectively, where $\mathbf{F}\in\mathbb{C}^{N\times N}$ and $\mathbf{G}\in\mathbb{C}^{M\times M}$ are full-rank mapping matrices (code-books). The estimated squared-loss mutual information based on a plug-in estimator is given by the Frobenius norm of a sample coherence matrix, that is:
\begin{equation}
||\hat{\mathbf{C}}||^{2}=\hat{I}_{s}\left(X;Y\right),\label{frobe}
\end{equation}
where 
\begin{equation}
\hat{\mathbf{C}}=\hat{\mathbf{R}}_{x}^{-1/2}\hat{\mathbf{C}}_{xy}\hat{\mathbf{R}}_{y}^{-1/2},\label{pseudoC}
\end{equation}
being $\hat{\mathbf{R}}_{x}=\mathbf{X}\mathbf{X}^{H}/L$ and $\hat{\mathbf{R}}_{y}=\mathbf{Y}\mathbf{Y}^{H}/L$ the sample autocorrelation matrices and $\hat{\mathbf{C}}_{xy}=\mathbf{X}\mathbf{P}_{\mathbf{1}}^{\bot}\mathbf{Y}^{H}/L$ the sample cross-covariance matrix. In particular, a sufficient condition for (\ref{frobe}) is that $\mathbf{F}=\mathbf{I}_{N}$ and $\mathbf{G}=\mathbf{I}_{M}$, which implies mapping the data onto the orthonormal canonical basis.
\end{thm}
\begin{IEEEproof}
See Appendix B.
\end{IEEEproof}
\vspace{8pt}Although Thm. \ref{theo coherence} sets the link between the SMI surrogate and second-order statistics, matrix $\hat{\mathbf{C}}$ in (\ref{pseudoC}) is not (apparently) a coherence matrix as that required by CCA, because autocorrelation instead of covariances are involved. However, the following theorem establishes the full link with CCA:
\begin{thm}
\label{theo coherence-2}Full link SMI-CCA: Let $\mathbf{X}\in\mathbb{C}^{N'\times L}$ ($N'<N$) and $\mathbf{Y}\in\mathbb{C}^{M'\times L}$ ($M'<M$) be data matrices obtained as $\mathbf{X}=\mathbf{F}\mathbf{D}_{x}$ and $\mathbf{Y}=\mathbf{G}\mathbf{D}_{y}$, respectively, where $\mathbf{F}\in\mathbb{C}^{N'\times N}$ and $\mathbf{G}\in\mathbb{C}^{M'\times M}$ are full-rank mapping matrices (code-books). Let us define the small-size sample coherence matrix as 
\begin{equation}
\hat{\mathbf{C}}_{N',M'}=\hat{\mathbf{C}}_{x}^{-1/2}\hat{\mathbf{C}}_{xy}\hat{\mathbf{C}}_{y}^{-1/2},\label{estasi}
\end{equation}
being $\hat{\mathbf{C}}_{x}=\mathbf{X}\mathbf{P}_{\mathbf{1}}^{\bot}\mathbf{X}^{H}/L$ and $\hat{\mathbf{C}}_{y}=\mathbf{Y}\mathbf{P}_{\mathbf{1}}^{\bot}\mathbf{Y}^{H}/L$ the sample covariance matrices and $\hat{\mathbf{C}}_{xy}=\mathbf{X}\mathbf{P}_{\mathbf{1}}^{\bot}\mathbf{Y}^{H}/L$ the sample cross-covariance matrix. Then:
\begin{equation}
||\hat{\mathbf{C}}_{N',M'}||^{2}\leq\hat{I}_{s}\left(X;Y\right).\label{ineq_cohes}
\end{equation}
In particular, a sufficient condition for the equality in (\ref{ineq_cohes}) is that $N'=N-1$, $M'=M-1$ and that the columns of $\mathbf{F}$ and $\mathbf{G}$ are given by the $(N-1)$-simplex and by the $(M-1)$-simplex, respectively.
\end{thm}
\textbf{Remark 1}. As a result of Thm. \ref{theo coherence-2}, we conclude that the coherence matrix required for estimating the SMI through its Frobenius norm can be estimated either by (\ref{pseudoC}) or (\ref{estasi}), provided that the Moore-Penrose inverse is generally used to cope with the rank-deficient case \cite{peze_2004}.
\begin{IEEEproof}
See Appendix C.
\end{IEEEproof}
\vspace{8pt}The physical meaning of matrices $\mathbf{F}$ and $\mathbf{G}$ in both Thms. \ref{theo coherence} and \ref{theo coherence-2} is that their columns contain the vectors to which the events of sources $X$ and $Y$ are mapped, respectively. The implication of Thm. \ref{theo coherence-2} is that, as $\hat{\mathbf{C}}_{N-1,M-1}$ in (\ref{estasi}) is just the sample coherence matrix required in CCA, the squared-loss mutual information can be expressed just as the sum of the squared canonical correlations:
\begin{equation}
\hat{I}_{s}\left(X;Y\right)=\sum_{i=1}^{\min(N,M)-1}\hat{\lambda}_{i}^{2}(\hat{\mathbf{C}}).\label{suma}
\end{equation}
Note also that, since a coherence matrix is invariant under linear invertible transforms, the code-books used for the SMI computation are irrelevant, provided that linearly independent vectors (columns of $\mathbf{F}$ and $\mathbf{G}$) are used. Otherwise, if the dimension of the space spanned after the mapping of $X$ is smaller than required (i.e. $N'<N-1$ and/or $M'<M-1$), the contribution of the smallest canonical correlations may be lost. The minimum dimension for the mapping of a source to vectors is therefore equal to the cardinality minus one. Moreover, the theorem also states implicitly that using higher dimension (i.e. $N'>N-1$ and/or $M'>M-1$) will yield a low-rank structure on $\hat{\mathbf{C}}_{x}$ and/or $\hat{\mathbf{C}}_{y}$. This idea will take a fundamental role in the process of leveraging all these notions to the analog case. In short, Fig. \ref{fig2} illustrates the stated notion behind Thm. \ref{theo coherence-2}: binary data can be mapped to $1$-dimensional points in the set $\{-1,1\}$; ternary data can be mapped to $2$-dimensional points in the set $\{[1,0],[-0.5,\sqrt{3}/2],[-0.5,-\sqrt{3}/2]\}$, and so on. 

\subsection{Relation to Gebelein Maximal Correlation}

Finally, the SMI measure can be linked with another important notion. Let $f:\mathcal{X}\rightarrow\mathbb{R}$ and $g:\mathcal{Y}\rightarrow\mathbb{R}$ be scalar representations of the sources $X$ and $Y$, respectively. For $\mathcal{X}=\mathcal{Y}=\mathbb{R}$, the Hirschfeld-Gebelein-R\'enyi (HGR) maximal correlation between sources $X$ and $Y$ is defined as \cite{renyi-1959,papadatos-2013}:
\begin{equation}
\rho(X;Y)=\sup_{\begin{array}{c}
f,g:\mathbb{E}_{p_{X}}f=\mathbb{E}_{p_{Y}}g=0,\\
\mathbb{E}_{p_{X}}f^{2}=\mathbb{E}_{p_{Y}}g^{2}=1
\end{array}}\mathbb{E}_{p_{XY}}[f(x)g(y)] =\sup_{f,g}\frac{\sigma_{fg}}{\sqrt{\sigma_{f}^{2}\sigma_{g}^{2}}},\label{supre}
\end{equation}
where 
\begin{equation*}
\sigma_{fg}=\mathbb{E}_{p_{XY}}[\left(f(x)-\mathbb{E}_{p_{X}}f(x)\right)\left(g(y)-\mathbb{E}_{p_{Y}}g(y)\right)],
\end{equation*}
\begin{equation*}
\sigma_{f}^{2}=\mathbb{E}_{p_{X}}[\left(f(x)-\mathbb{E}_{p_{X}}f(x)\right)^{2}],
\end{equation*}
\begin{equation}
\sigma_{g}^{2}=\mathbb{E}_{p_{Y}}[\left(g(y)-\mathbb{E}_{p_{Y}}g(y)\right)^{2}],
\end{equation}
and the supremum in (\ref{supre}) is taken over all Borel functions $f$ and $g$. The HGR maximal correlation $\rho(X;Y)$ represents the maximal Pearson coefficient that can be obtained after mapping the events of the sources to reals, and it has found numerous applications in information theory and statistics (see \cite{Anantharam-2013} and reference therein). Recently, the HGR has been proposed as a practical and more relevant surrogate of mutual information in the field of security and privacy \cite{ting-2018}. 

\begin{figure}[tp]
\centering
\includegraphics[clip,scale=0.6]{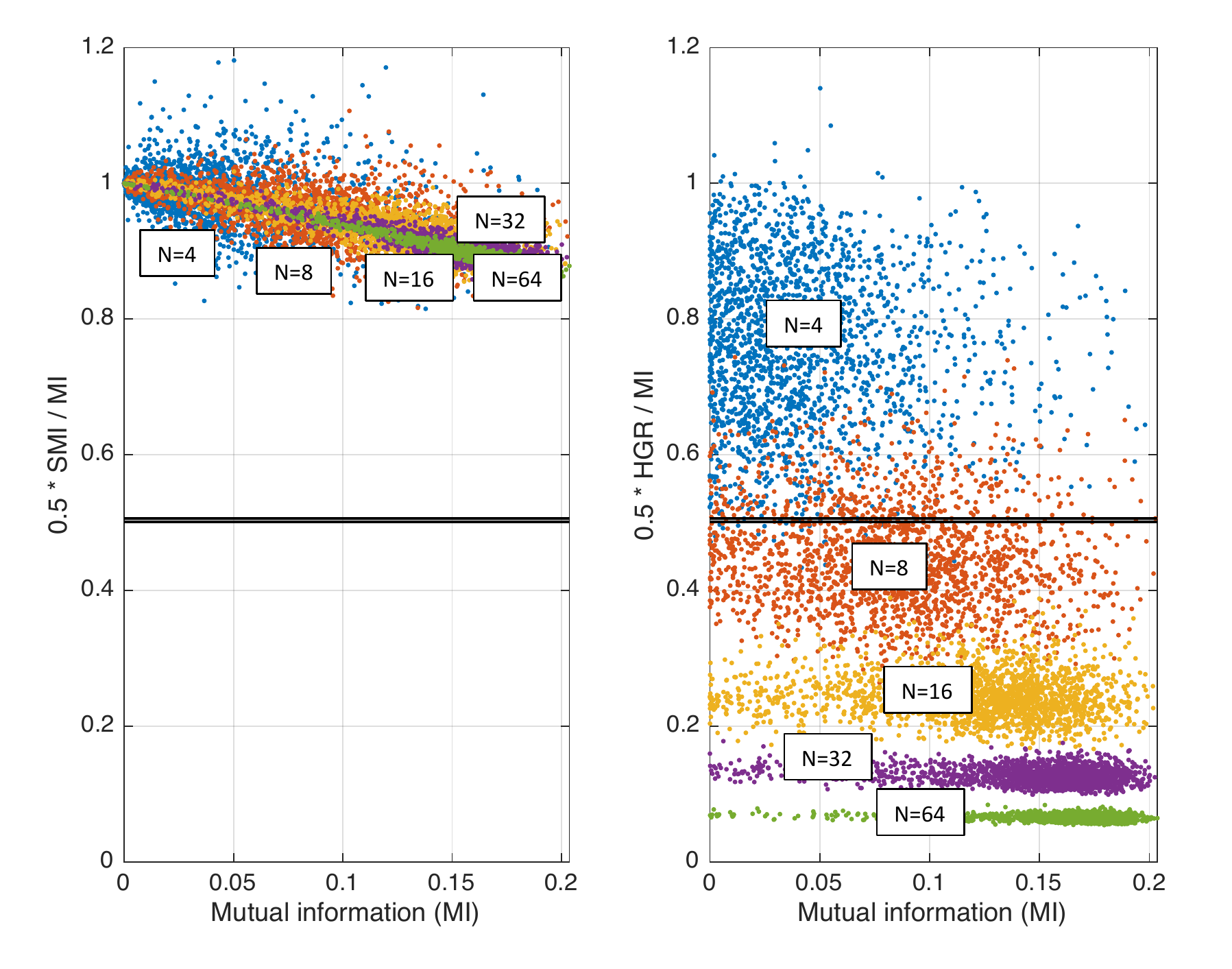}\caption{Half of the ratio between the squared-loss MI and MI (left) and between the squared HGR and MI (right) for random discrete memory-less channels with random input distributions, for different $N$ ($=M$) values of alphabet sizes. \label{fig3}}
\end{figure}

For sources with finite alphabets $\mathcal{X}$ and $\mathcal{Y}$, the problem of estimating the HGR maximal correlation coefficient can be reformulated as follows. Let $\mathbf{u}\in\mathbb{R}^{N}$ and $\mathbf{v}\in\mathbb{R}^{M}$ be the vectors containing the reals towards which the events of sources $X$ and $Y$ are mapped, respectively, that is $[\mathbf{u}]_{n}=f(x_{n})$ for $n=1:N$ and $[\mathbf{v}]_{m}=g(y_{m})$ for $m=1:M$. Then, from a sequence of $L$ i.i.d. pairs $\{x(l),y(l)\}$ we obtain the pairs of $L$-th length samples $\{\mathbf{u}^{H}\mathbf{D}_{x},\mathbf{v}^{H}\mathbf{D}_{y}\}$. Clearly,
\begin{equation}
\hat{\rho}(X;Y)=\max_{\mathbf{u},\mathbf{v}}\frac{\mathbf{u}^{H}\hat{\mathbf{C}}_{xy}\mathbf{v}}{\sqrt{\mathbf{u}^{H}\hat{\mathbf{C}}_{x}\mathbf{u}}\sqrt{\mathbf{v}^{H}\hat{\mathbf{C}}_{y}\mathbf{v}}},
\end{equation}
which is given by the maximum singular value of the empirical coherence matrix $\hat{\mathbf{C}}$, implying that $0\leq\hat{\rho}(X;Y)\leq1$. In contrast, according to what we have shown in this paper, the SMI is given by the sum of the squares of all potentially non-zero singular values of the coherence matrix, as seen in (\ref{suma}). Therefore, apart from the \textit{best} single mapping to reals that the HGR notion provides, the SMI looks as well to other mappings to canonical coordinates of the coherence, thus becoming more sensitive to complex hidden relationships between the observed data. To illustrate these ideas, Fig. \ref{fig3} compares the two measures of information in what concerns to their relation with mutual information by testing discrete memory-less channels generated randomly with input distribution also generated randomly. It is seen that, in contrast with the HGR measure, the SMI exhibits a consistent behavior at the small dependence regime, in the sense of its much smaller dispersion around the value of 1 of the ratio between half of the HGR measure and the true MI, as well as its much less sensitivity to the alphabet size of the sources. It should be noted that, under the presented unified vision, the HGR maximal correlation coefficient can be seen as an extreme case of (badly) measuring the SMI by mapping the sources to reals (or a dimension spanned by the mapping equal to 1) instead of mapping them onto linearly independent vectors or onto the simplex, and it represents the maximum loss of information with respect to the true MI. 

\section{Analog sources: second-order statistics on the characteristic feature space}

For discrete sources, we have shown that estimating the SMI surrogate of mutual information via second-order statistics entails the mapping of events onto a vectorial space spanning a minimum dimension equal to the source cardinality minus one. But analog sources are of infinite dimension, so an infinite dimensional mapping is in principle required to retain all the information. This key idea, informally stated in Cover's theorem on the separability of patterns \cite{haykinbook-2009}\footnote{``\textit{A complex pattern-classification problem, cast in a high-dimensional space nonlinearly, is more likely to be linearly separable than in a low-dimensional space, provided that the space is not densely populated}'' (T. M. Cover).}, is well known in the field of machine learning. In particular, kernel methods have the ability (called \textit{kernel trick} \cite{martinez-2018}) of implicitly using linear algebra on high (\textit{infinite}) dimensional spaces without the necessity of explicitly visiting that huge space.

Leaving the kernel paradigm, in this section we propose an explicit mapping of analog sources onto a space of \textit{finite} dimension on the complex field. The motivation is two-fold: on the one hand, the explicit mapping allows the use of the standard CCA as seen for discrete sources; on the other hand, the fixed dimension acts itself as a regularizer of the problem. The advantage is that, while the computation of scalar products on the implicit high dimensional spaces is very direct by using kernel methods, it is not so clear how to implement the inversion of matrices as those required by CCA. Although \textit{kernelized} versions of CCA (KCCA) have been proposed (see \cite{bach-2002,hardoon-2003}) in the context of several signal processing and machine learning applications (e.g. blind source separation and nonlinear channel identification/equalization \cite{vaere-2013}), they involve costly inversion of big Gram matrices of kernel dot products between all data pairs, thus requiring strategies for decreasing complexity, such as the \textit{incomplete Cholesky factorization }\cite{martinez-2018}. In addition, kernel methods ultimately need to be regularized to avoid overfitting. In that sense, the alternative based on an explicit mapping proposed in the sequel can be seen as procedure for regularizing the problem from the beginning, providing interpretability and computational complexity savings.


\subsection{Dependence, correlation and characteristic function}

To motivate the mapping that will be finally proposed in (\ref{maping_cha}), let us write the marginal and joint characteristic functions (CF) (defined as the Fourier transform of the PDFs with sign reversal in the complex exponential) of a pair of analog sources $X$ and $Y$ as follows:
\begin{equation*}
\varphi_{X}(\omega_{1})=\intop p_{X}(x)e^{j\omega_{1}x}dx=\mathbb{E}_{p_{X}}\left[z_{1}\right],
\end{equation*}
\begin{equation*}
\varphi_{Y}(\omega_{2})=\intop p_{Y}(y)e^{j\omega_{2}y}dy=\mathbb{E}_{p_{Y}}\left[z_{2}\right],
\end{equation*}
\begin{equation}
\varphi_{XY}(\omega_{1},\omega_{2})=\mkern-5mu\intop\mkern-5mup_{XY}(x,y)e^{j(\omega_{1}x+\omega_{2}y)}dxdy=\mathbb{E}_{p_{XY}}\left[z_{1}z_{2}\right],
\end{equation}
where $z_{1}=e^{j\omega_{1}X}$ and $z_{2}=e^{j\omega_{2}Y}$ are complex random variables obtained from $X$ and $Y$ through a nonlinear mapping. Clearly, if $X$ and $Y$ are independent, then $\varphi_{XY}(\omega_{1},\omega_{2})=\mathbb{E}_{p_{X}}\left[z_{1}\right]\mathbb{E}_{p_{Y}}\left[z_{2}\right]=\varphi_{X}(\omega_{1})\varphi_{Y}(\omega_{2})$ for all possible values of $\omega_{1}$ and $\omega_{2}$, implying that $z_{1}$ and $z_{2}$ are uncorrelated random variables. Note that the converse is also true: if $z_{1}$ and $z_{2}$ are uncorrelated for all possible values of $\omega_{1}$ and $\omega_{2}$, then $X$ and $Y$ are independent, since the condition $\varphi_{XY}(\omega_{1},\omega_{2})=\varphi_{X}(\omega_{1})\varphi_{Y}(\omega_{2})$ implies that $p_{XY}(x,y)=p_{X}(x)p_{Y}(y)$ as a result of the bijective property of the Fourier transform. It is important to emphasize that the converse statement mentioned above guarantees that any kind of statistical dependence between $X$ and $Y$ will be ``manifested'' as correlation for some values of $\omega_{1}$ and $\omega_{2}$, which means that the set of complex exponential functions is not restrictive for the problem of independence detection via second-order statistics. In other words, independence detection can be formulated as a problem of correlation detection (see \cite{cabrera-2018}) by resorting to the characteristic function, provided that sufficient number of values of $\omega_{1}$ and $\omega_{2}$ are explored. 

From this observation, two natural questions arise for the problem of SMI estimation: how many points of $\omega_{1}$ and $\omega_{2}$ for correlation analysis need to be explored? How small the separation between the explored points needs to be? To answer to these questions, we next propose a finite support for regularization (Sec. \ref{subsec:Regularization-through-Gaussian}) and a uniform sampling (Secs. \ref{subsec:Regularization-through-Gaussian}\&\ref{subsec:Second-order-statistics-on}) of the characteristic function, which further yield to an efficient estimation approach (Sec. \ref{subsec:Large-feature-space}).


\subsection{Regularization through Gaussian convolutions\label{subsec:Regularization-through-Gaussian}}

It is well known that the problem of estimating differential entropy and mutual information needs to be regularized \cite{wang_2009}. In the sequel, we propose a regularization approach based on the properties of the characteristic function. The core idea is the concept of Gaussian convolutions, which has been recently proposed in \cite{goldfeld_2019} in the framework of differential entropy estimation as a means to achieve the parametric rate of convergence (w.r.t. the sample size) for distributions belonging to any nonparametric class. In the context of this paper, the approach has the additional advantage of providing a clear physical meaning to the proposed estimators of information, as seen in the sequel. 

Consider that sources $X$ and $Y$ are in fact contaminated by independent zero-mean additive Gaussian sources $V_{x}$ and $V_{y}$ with known smoothing variance $\sigma^{2}$ and PDF $p_{V}$:
\begin{equation}
x'(l)=x(l)+v_{x}(l),\,\,\,\,\,y'(l)=y(l)+v_{y}(l).
\end{equation}
The purpose is now to estimate the contaminated information between the virtual sources $x'(l)$ and $y'(l)$ using the data obtained from the actual sources $x(l)$ and $y(l)$, which are still accessible. By doing so, a natural regularization of the problem is achieved, as seen in the sequel. 

Since the PDF of the sum of independent random variables is the convolution of densities, that is $p_{X'}(x)=p_{X}(x)*p_{V}(x)$ and $p_{Y'}(y)=p_{Y}(y)*p_{V}(y)$, the CF is just the product of CFs of each term, that is $\varphi_{X'}(\omega)=\varphi_{X}(\omega)\varphi_{V}(\omega)$ and $\varphi_{X'}(\omega)=\varphi_{Y}(\omega)\varphi_{V}(\omega)$, where 
\begin{equation}
\varphi_{V}(\omega)=e^{-\sigma^{2}w^{2}/2}\label{window}
\end{equation}
is the CF of both $V_{x}$ and $V_{y}$. The key point is that the Gaussian shape has an effective support, which allows focusing on a finite interval given by $|\omega|\leq\omega_{\max}=k\sigma^{-1}$, typically with $k=2.5$. Consequently, as $|\varphi_{X}(\omega)|\leq1$ and $|\varphi_{Y}(\omega)|\leq1$, the CFs of the contaminated sources $X'$ and $Y'$ become both roughly zero for $|\omega|>\omega_{\max}$ as well. The higher is $\sigma^{2}$, the stronger is the smoothing effect caused on the PDFs, and the smaller is the effective support of the CFs, exhibiting an insightful duality with the classical spectral estimation problem. Note that, by the general data processing inequality for $f$-divergences (see \cite{collet-2019} and references therein), the additive perturbation in both sources regularizes the problem by decreasing and bounding the amount of mutual information to be measured, yielding to a negative bias contribution to the estimators as verified later on with computer simulations. In summary, estimates of the CFs of the contaminated sources can be obtained by just tapering the sample mean estimators as follows:
\begin{equation}
\hat{\varphi}_{X'}(\omega)=\left\langle e^{j\omega x(l)}\right\rangle _{L}\varphi_{V}(\omega),\,\,\,\,\hat{\varphi}_{Y'}(\omega)=\left\langle e^{j\omega y(l)}\right\rangle _{L}\varphi_{V}(\omega).\label{cha weighting}
\end{equation}

Once the effective support of the empirical CFs is fixed, consider a uniform sampling in the $\omega$ domain with sampling period $\alpha$. As CFs and PDFs are Fourier pairs, the sampling of CFs implies a periodic extension of the PDFs, such that the implicit density of $X$ becomes
\begin{equation}
p_{X'}(x)=\sum_{k}(p_{X}*p_{V})\left(x-\frac{2\pi k}{\alpha}\right),\label{periodic}
\end{equation}
and similarly for $Y$. The smaller is $\alpha$, the smaller is the aliasing effect in (\ref{periodic}), so the sampling period $\alpha$ can be roughly determined as the inverse of the expected dynamic range of the PDFs of the sources, that is $\alpha=1/(q\sigma_{x})$, typically with $q=3$. Assuming a CF support of $2\omega_{\max}$, this yields a number of sampling points of the CFs given by 
\begin{equation}
N=2\left\lceil \frac{\omega_{\max}}{\alpha}\right\rceil +1=2\left\lceil kq\frac{\sigma_{x}}{\sigma}\right\rceil +1,\label{valor de N}
\end{equation}
where an odd value of $N$ is imposed for clarity in forthcoming developments. It is worth noting that, as the Gaussian window minimizes the uncertainty principle, a commonly used rationale in the classical time-frequency analysis, an additive Gaussian perturbation will therefore minimize the effective support of the contaminated characteristic function (i.e. the effective dimension of the feature space) for a given smoothing variance $\sigma^{2}$, which further supports the rationale for using the tool of Gaussian convolutions as a natural regularizer in the specific methodology explored in this paper. The interpretation of (\ref{valor de N}) is that of moving the problem to a finite parametric representation of the PDFs, which originally belong to a nonparametric class. Then, as the implicit number of parameters of the problem becomes finite, the SMI estimation problem will turn out to be consistent. 


\subsection{Second-order statistics on the characteristic space and SMI estimate\label{subsec:Second-order-statistics-on}}

Given the physical sense of the proposed regularization, we propose (see \cite{cabrera-2019}) a uniform symmetric and finite sampling of $\omega_{1}$ and $\omega_{2}$ to define mappings $\phi_{X}(.):\mathbb{R}\rightarrow\mathbb{C}^{N}$ and $\phi_{Y}(.):\mathbb{R}\rightarrow\mathbb{C}^{N}$ as
\begin{equation}
x\rightarrow\mathbf{x}=e^{j\alpha\mathbf{n}x}\qquad y\rightarrow\mathbf{y}=e^{j\alpha\mathbf{n}y},\label{maping_cha}
\end{equation}
respectively, where $\mathbf{n}\in\mathbb{Z}^{N\times1}$ is a vector of integers defined as $\mathbf{n}=[-K,-K+1,\cdots,K]^{T}$ with $N=2K+1$. To appreciate the rationale, note that if one lets $\alpha\rightarrow0$ and $N\rightarrow\infty$ simultaneously in such a way that $N\alpha\rightarrow\infty$ as well, for instance $\alpha=O(N^{-1/2})$, we are then mapping the sources onto asymptotically orthogonal vectors, which ensures that the SMI estimate that we developed for discrete sources (based now on CCA performed on the new spaces) will be asymptotically unbiased, according with Thms. \ref{theo coherence} and \ref{theo coherence-2}. Note that the required feature space dimension is determined by (\ref{valor de N}), which explains why using a finite dimension acts as a natural regularization of the problem.

Consider a sequence of $L$ i.i.d. pairs $\{x(l),y(l)\}\in\mathbb{R}^{2}$ for $l=1,2,\ldots,L$. Using the mappings defined in (\ref{maping_cha}), we obtain the pair of vector sequences $\{\mathbf{x}(l),\mathbf{y}(l)\}\in\mathbb{\mathbb{C}}^{N\times2}$ in the feature space and construct the data matrices $\mathbf{X}\in\mathbb{C}^{N\times L}$ and $\mathbf{Y}\in\mathbb{C}^{N\times L}$ as follows:
\begin{equation}
[\mathbf{X}]_{:,l}=\mathbf{x}(l),\qquad[\mathbf{Y}]_{:,l}=\mathbf{y}(l).
\end{equation}
From Thm. \ref{theo coherence} and Remark 1, the SMI for analog sources can be finally estimated as 
\begin{equation}
\hat{I}_{s}\left(X;Y\right)=||\hat{\mathbf{C}}||^{2},\label{for analog}
\end{equation}
with 
\begin{equation}
\hat{\mathbf{C}}=\hat{\mathbf{R}}_{x'}^{-1/2}\hat{\mathbf{C}}_{x'y'}\hat{\mathbf{R}}_{y'}^{-1/2}.\label{la ce analog}
\end{equation}
Note that, following the concept of Gaussian convolutions, the sample autocorrelations and the cross-covariance in (\ref{la ce analog}) refer to the contaminated sources $X'$ and $Y'$, for which the result in (\ref{cha weighting}) is used in the sequel to compute both the cross-covariance and the autocorrelation matrices.

On the one hand, concerning the cross-covariance in (\ref{la ce analog}), we clearly obtain from (\ref{cha weighting}) that:
\begin{equation}
\hat{\mathbf{C}}_{x'y'}=\left\langle e^{j\alpha\mathbf{n}x(l)}e^{-j\alpha\mathbf{n}^{T}y(l)}\right\rangle _{L}\odot\left(\mathbf{w}\mathbf{w}^{T}\right)-\hat{\mathbf{p}}\hat{\mathbf{q}}^{H}
\end{equation}
where the weighted first-order statistics are
\begin{equation}
\hat{\mathbf{p}}=\left\langle e^{j\alpha\mathbf{n}x(l)}\right\rangle _{L}\odot\mathbf{w},\,\,\,\,\,\hat{\mathbf{q}}=\left\langle e^{j\alpha\mathbf{n}y(l)}\right\rangle _{L}\odot\mathbf{w}
\end{equation}
and the symmetric tapering vector is defined as
\begin{equation}
[\mathbf{w}]_{n}=\varphi_{V}((n-K)\alpha)=e^{-\sigma^{2}\alpha^{2}(n-K)^{2}/2}\label{taptering}
\end{equation}
for $n=0,1,\ldots,N-1$.

On the other hand, the elements of the sample autocorrelation matrices in (\ref{la ce analog}) can be expressed as $[\hat{\mathbf{R}}_{x}]_{n,m}=\left\langle e^{j\alpha(n-m)x(l)}\right\rangle _{L}\varphi_{V}(\alpha(n-m))$ and $[\hat{\mathbf{R}}_{y}]_{n,m}=\left\langle e^{j\alpha(n-m)y(l)}\right\rangle _{L}\varphi_{V}(\alpha(n-m))$ for $n,m=0,1,\ldots,2K$, which endows them with a Toeplitz structure. As a result, we can construct them as follows:
\begin{equation}
\hat{\mathbf{R}}_{x'}=\text{Toe}\left(\hat{\mathbf{p}}_{a}\right),\,\,\,\,\,\hat{\mathbf{R}}_{y'}=\text{Toe}\left(\hat{\mathbf{q}}_{a}\right),
\end{equation}
where $\hat{\mathbf{p}}_{a}$ and $\hat{\mathbf{q}}_{a}$ are defined as the extended weighted first-order statistics, 
\begin{equation}
\hat{\mathbf{p}}_{a}=\left\langle e^{j\alpha\mathbf{n}_{a}x(l)}\right\rangle _{L}\odot\mathbf{w}_{a},\,\,\,\,\,\hat{\mathbf{q}}_{a}=\left\langle e^{j\alpha\mathbf{n}_{a}y(l)}\right\rangle _{L}\odot\mathbf{w}_{a},\label{tap1}
\end{equation}
with $\mathbf{n}_{a}=[0,1,\cdots,N-1]^{T}$ and the asymmetric tapering vector is defined as 
\begin{equation}
[\mathbf{w}_{a}]_{n}=\varphi_{V}(n\alpha)=e^{-\sigma^{2}\alpha^{2}n^{2}/2}\label{tap2}
\end{equation}
for $n=0,1,\ldots,N-1$.

As a final remark, note that the regularization technique proposed above differs from the classical regularization technique used in KCCA based on diagonal loading of autocorrelation matrices \cite{bach-2002}. Although both techniques succeed in solving the rank deficient issue, the proposed regularization based on tapering provides physical interpretation to the overall effect on the final estimate. 


\subsection{Large feature space dimension regime approximation\label{subsec:Large-feature-space}}

The Toeplitz structure of $\hat{\mathbf{R}}_{x'}$ and $\hat{\mathbf{R}}_{y'}$ can be further exploited for the computation of the inverses in (\ref{la ce analog}). Szeg\"o's theorem (see \cite{grenander-1958,gray-2006}) establishes that, for large dimension, a Toeplitz matrix is asymptotically diagonalizable by the unitary Fourier matrix, and its eigenvalues asymptotically
behave like samples of the Fourier transform of its entries. The most general and relaxed assumption that guarantees the behavior stated in Szeg\"os theorem is that the columns of the matrices are square-integrable for $N\rightarrow\infty$. This condition is clearly ensured by the tapering operation in (\ref{tap1})\&(\ref{tap2}). Effectively, as the Gaussian taper in (\ref{tap2}) is square-integrable for any $\sigma^{2}>0$ and the sample CFs are upper-bounded, that is $|\left\langle e^{j\alpha nx(l)}\right\rangle _{L}|\leq1$ and $|\left\langle e^{j\alpha ny(l)}\right\rangle _{L}|\leq1$, then the sample vectors $\hat{\mathbf{p}}_{a}$ and $\hat{\mathbf{q}}_{a}$ become square-integrable for $N\rightarrow\infty$. This fact motivates a frequency-domain tool to reduce complexity by leveraging the approximate diagonalization of the involved Toeplitz matrices after a Fourier transform. In particular, the following theorem sets the required theoretical framework:
\begin{thm}
\label{thm szego}Let $t_{n}\in\mathbb{C}$ be an Hermitian sequence such that $t_{0}=1$ and $\lim_{N\rightarrow\infty}\sum_{n=0}^{N-1}|t_{n}|^{2}<\infty$. Let us define vector $\mathbf{t}\in\mathbb{C}^{N}$ and Hermitian-Toeplitz matrix $\mathbf{T}\in\mathbb{C}^{N\times N}$ as $[\mathbf{t}]_{n}=t_{n}$ and $\mathbf{T}=\mathrm{\textrm{\text{Toe}}}\left(\mathbf{t}\right)$, respectively. Let $\text{\ensuremath{\mathbf{U}}}\in\mathbb{C}^{N\times N}$ be the unitary Fourier matrix. Then
\begin{equation}
\lim_{N\rightarrow\infty}\left\{ \left[\text{\ensuremath{\mathbf{U}}}\mathbf{T}\mathbf{U}^{H}\right]_{n,m}-\right.\left.\left(2\sqrt{N}\textrm{\text{Re}}\left(\left[\mathbf{U}^{H}\left(\mathbf{t}\odot\mathbf{v}\right)\right]_{n}\right)-1\right)\delta_{nm}\right\} =0\label{eq th szego}
\end{equation}
for $n,m=0,1,\ldots,N-1$, where $\mathbf{v}$ is a unilateral triangular window with elements $[\mathbf{v}]_{n}=1-n/N$.
\end{thm}
\begin{IEEEproof}
See \cite{gray-2006} for detailed proofs concerning the limit behavior. In addition, in (\ref{eq th szego}) we have used that the Fourier transform of an Hermitian sequence $g_{n}$ can be written as $\sum_{n=-(N-1)}^{(N-1)}g_{n}e^{-j2\pi wn}=g_{0}+2\text{Re}\left(\sum_{n=1}^{(N-1)}g_{n}e^{-j2\pi wn}\right)=2\text{Re}\left(\sum_{n=0}^{(N-1)}g_{n}e^{-j2\pi wn}\right)-g_{0}$, and that $g_{n}=t_{n}(1-n/N)$ with $g_{0}=1$.
\end{IEEEproof}
Consider now the SMI estimate proposed in (\ref{for analog})\&(\ref{la ce analog}), which can be expressed as follows:
\begin{equation}
\hat{I}_{s}\left(X;Y\right)=\left\Vert \hat{\mathbf{R}}_{x'}^{-1/2}\hat{\mathbf{C}}_{x'y'}\hat{\mathbf{R}}_{y'}^{-1/2}\right\Vert ^{2}.
\end{equation}
As the Frobenious norm is invariant under unitary transforms, we can write
\begin{equation}
\hat{I}_{s}\left(X;Y\right)= \left\Vert (\mathbf{U}\hat{\mathbf{R}}_{x'}\mathbf{U}^{H})^{-1/2}\mathbf{U}\hat{\mathbf{C}}_{x'y'}\mathbf{U}^{H}(\mathbf{U}\hat{\mathbf{R}}_{y'}\mathbf{U}^{H})^{-1/2}\right\Vert ^{2}.\label{vista asi-1}
\end{equation}
Thm. \ref{thm szego} states that the transformed autocorrelation matrices $\mathbf{U}\hat{\mathbf{R}}_{x'}\mathbf{U}^{H}$ and $\mathbf{U}\hat{\mathbf{R}}_{y'}\mathbf{U}^{H}$ in (\ref{vista asi-1}) are asymptotically diagonal, which allows the formulation of an approximate and computationally efficient SMI estimator for the large dimension regime in the following manner:
\begin{equation}
\hat{I}_{as}\left(X;Y\right)=\left\Vert [\hat{\mathbf{p}}']^{-1/2}\mathbf{U}\hat{\mathbf{C}}_{x'y'}\mathbf{U}^{H}[\hat{\mathbf{q}}']^{-1/2}\right\Vert ^{2},\label{asymp}
\end{equation}
where, as $[\hat{\mathbf{p}}_{a}]_{0}=[\hat{\mathbf{q}}_{a}]_{0}=1$, we can use (\ref{tap1})\&(\ref{eq th szego}) to write the final transformed vectors as:
\begin{equation*}
\hat{\mathbf{p}}'=2\sqrt{N}\text{Re}\left(\mathbf{U}^{H}\left(\hat{\mathbf{p}}_{a}\odot\mathbf{v}\right)\right)-\mathbf{1},
\end{equation*}
\begin{equation}
\hat{\mathbf{q}}'=2\sqrt{N}\text{Re}\left(\mathbf{U}^{H}\left(\hat{\mathbf{q}}_{a}\odot\mathbf{v}\right)\right)-\mathbf{1}.
\end{equation}
The main advantage of the proposed approximate estimator in (\ref{asymp}) is that matrix inverses are avoided and only element-wise inverses are required. 


\section{Numerical results}

\begin{figure}[tp]
\centering
\includegraphics[scale=0.6]{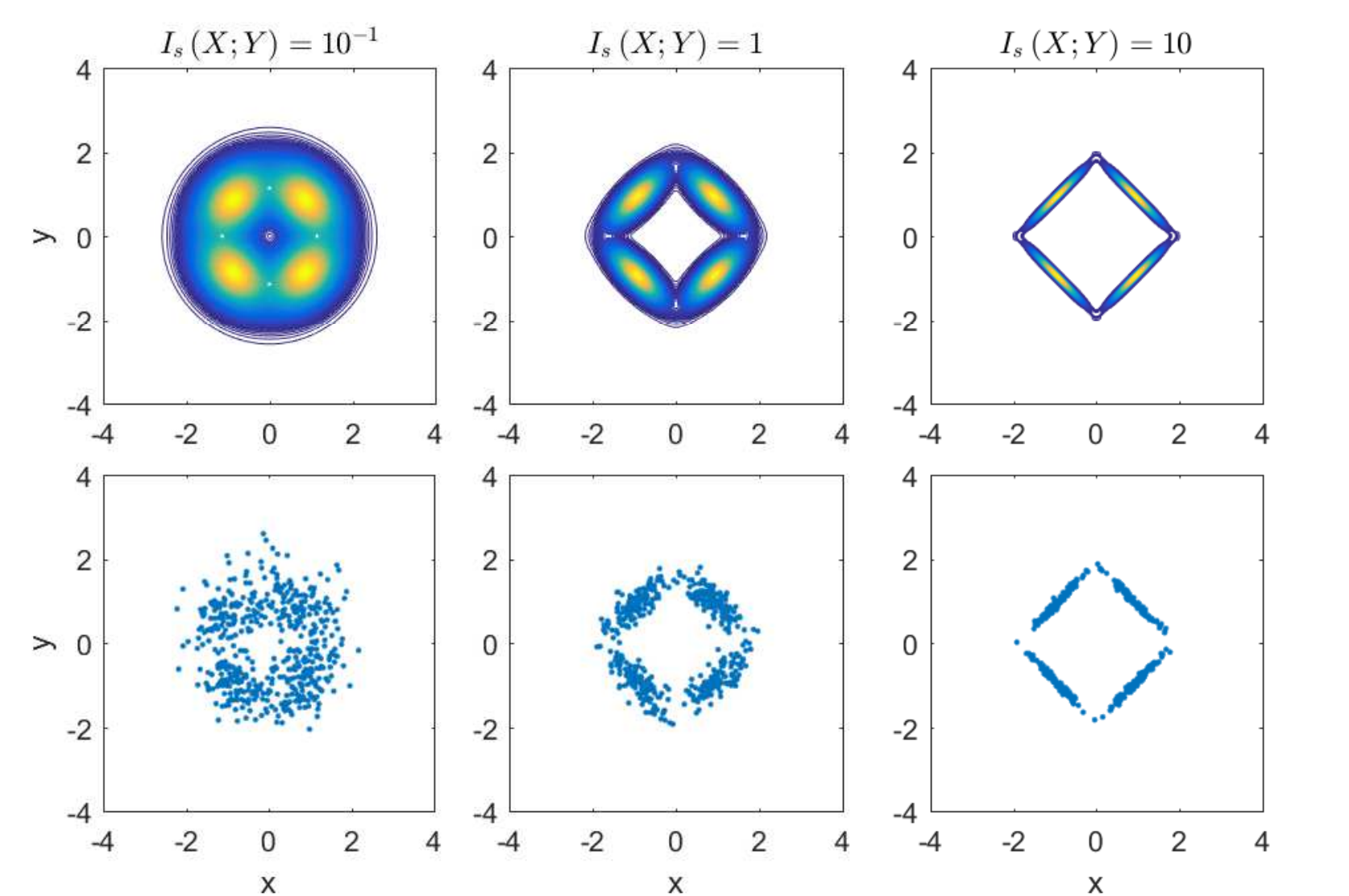}\caption{Examples of contour plots (up) and raw data (down) for small, medium and high dependence (left to right), all with null correlation. \label{fig4}}
\end{figure}
The performance of the proposed estimators, and the impact of their free parameters, is evaluated by means of Monte Carlo simulations. We measure the mean and variance of the estimated amount of information using the Gaussian Mixture Model (GMM) proposed in \cite{cabrera-2018}, which is illustrated in Fig. \ref{fig4}. The data is normalized such that $\mathbb{E}_{p_{X}}[x]=\mathbb{E}_{p_{Y}}[y]=0$, $\mathbb{E}_{p_{X}}[x^{2}]=\mathbb{E}_{p_{Y}}[y^{2}]=1$. The usefulness of this model lies on the fact that $\mathbb{E}_{p_{XY}}[xy]=0$ for any value of MI, thus forcing the estimators to discover dependence from uncorrelated data. To test the estimators, the true value of $I_{s}(X;Y)$ is obtained by a genie-aided estimator based on empirical averaging \cite{wang_2009} under the knowledge of $p_{XY}$. 
\begin{figure}[tp]
\centering
\includegraphics[clip,scale=0.6]{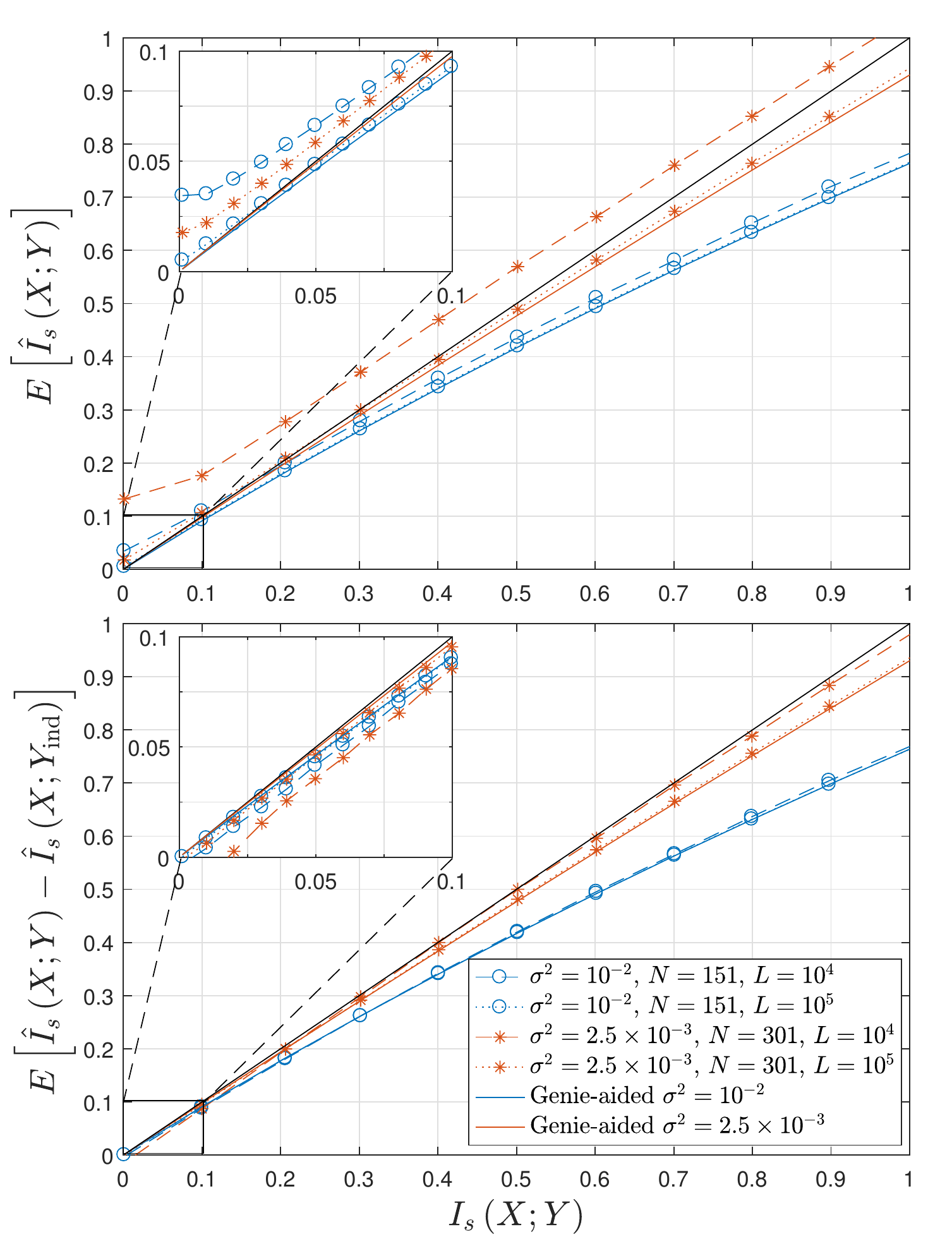}\caption{Mean of the estimated SMI (up) and reduced-bias estimators (down) as a function of the true SMI for $\alpha=1/3$, showing the role of $\sigma^{2}$ and $L$, with $N=2\left\lceil 7.5/\sigma\right\rceil +1$.
\label{fig5}}
\end{figure}

Fig. \ref{fig5} shows the mean of the proposed SMI estimators as a function of small ($I_{s}\left(X,Y\right)\in[0,0.1]$) and moderate ($I_{s}\left(X,Y\right)\in(0.1,1]$) values of true SMI, for different values of the dimension-variance pair $(\sigma^{2},N)$ and data size $L$. The feature space dimension $N$ is fixed from $\sigma^{2}$ by (\ref{valor de N}) using $k=3$ and $q=2.5$. Clearly, the linearity range of all curves increases as $\sigma^{2}$ decreases (and $N$ increases accordingly) in the large SMI regime by increasing the ceiling value, with the price of additionally increasing the SMI floor at the small SMI regime. For a given pair $(\sigma^{2},N)$, that floor gets inversely proportional to $L$. In short, the small dependence regime is the data limited regime, and the strong dependence regime is the dimension limited regime. In order to compensate the floor level at small data regime, a reduced bias estimator $\hat{I}_{s}\left(X,Y_{\text{}}\right)-\hat{I}_{s}\left(X,Y_{\text{ind}}\right)$ is also shown, where $Y_{\text{ind}}$ is independent data identically distributed as $Y$ and obtained by circularly shifting the data sequence associated to $Y$ by $j$ positions with $j\neq0$ and $j\neq L$. In this way, the residual biases associated to estimates of theoretically null squared canonical correlations are reduced, thus improving the impact on the overall bias at the small data regime for a sufficiently small (big) smoothing variance $\sigma^{2}$ (dimension $N$), and regardless of the kind of data statistics. Finally, to validate the regularization based on Gaussian convolutions, another genie-aided estimate computed from truly contaminated data with independent additive Gaussian noises of variance $\sigma^{2}$ is also shown. As expected, the proposed estimators become asymptotically close from above to the contaminated SMI value as $L\rightarrow\infty$, which provides a physical interpretation of the negative bias that emerges at the moderate dependence regime.

\begin{figure}
\subfloat{\includegraphics[scale=0.6]{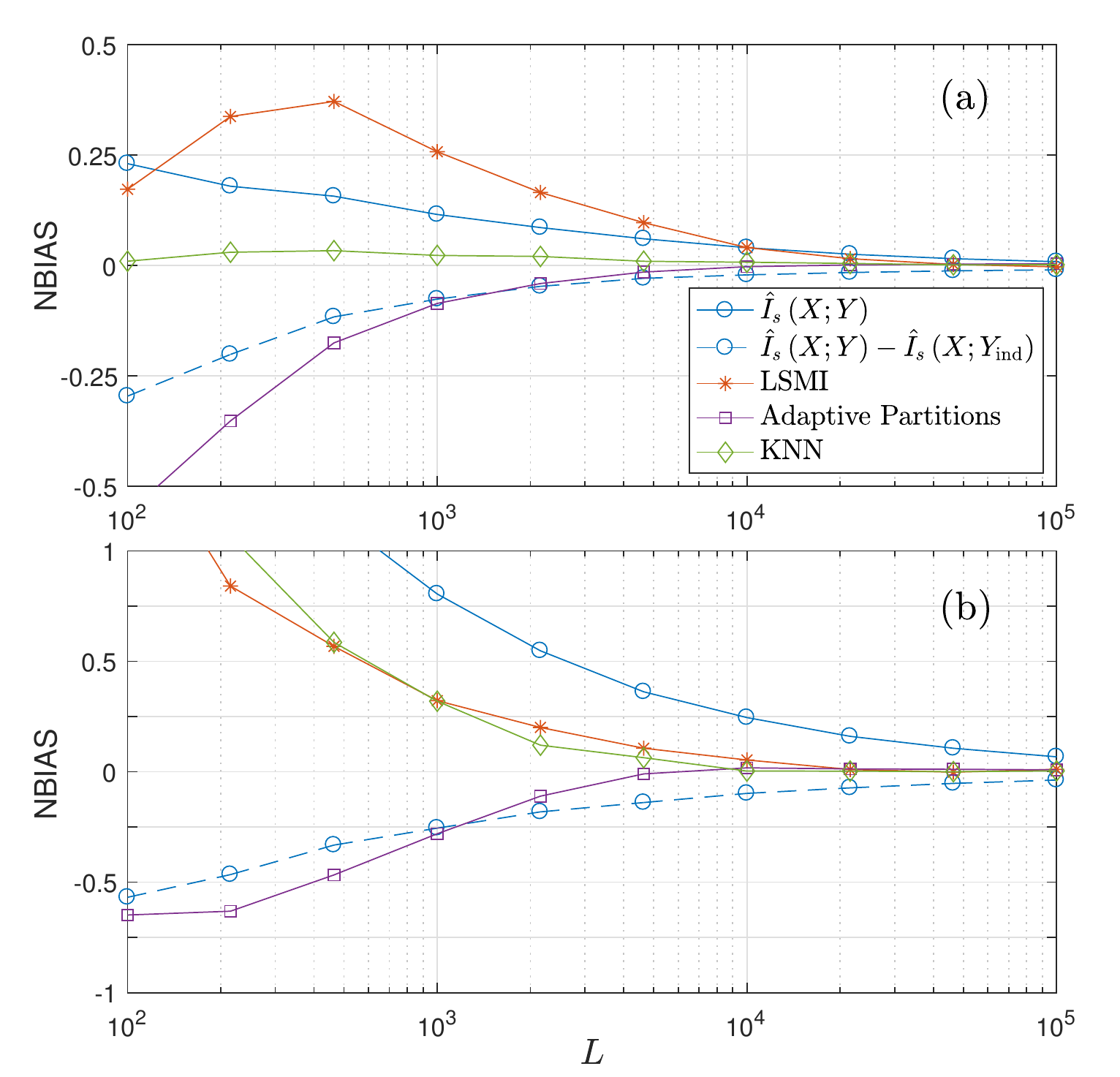}} 
\subfloat{\includegraphics[scale=0.6]{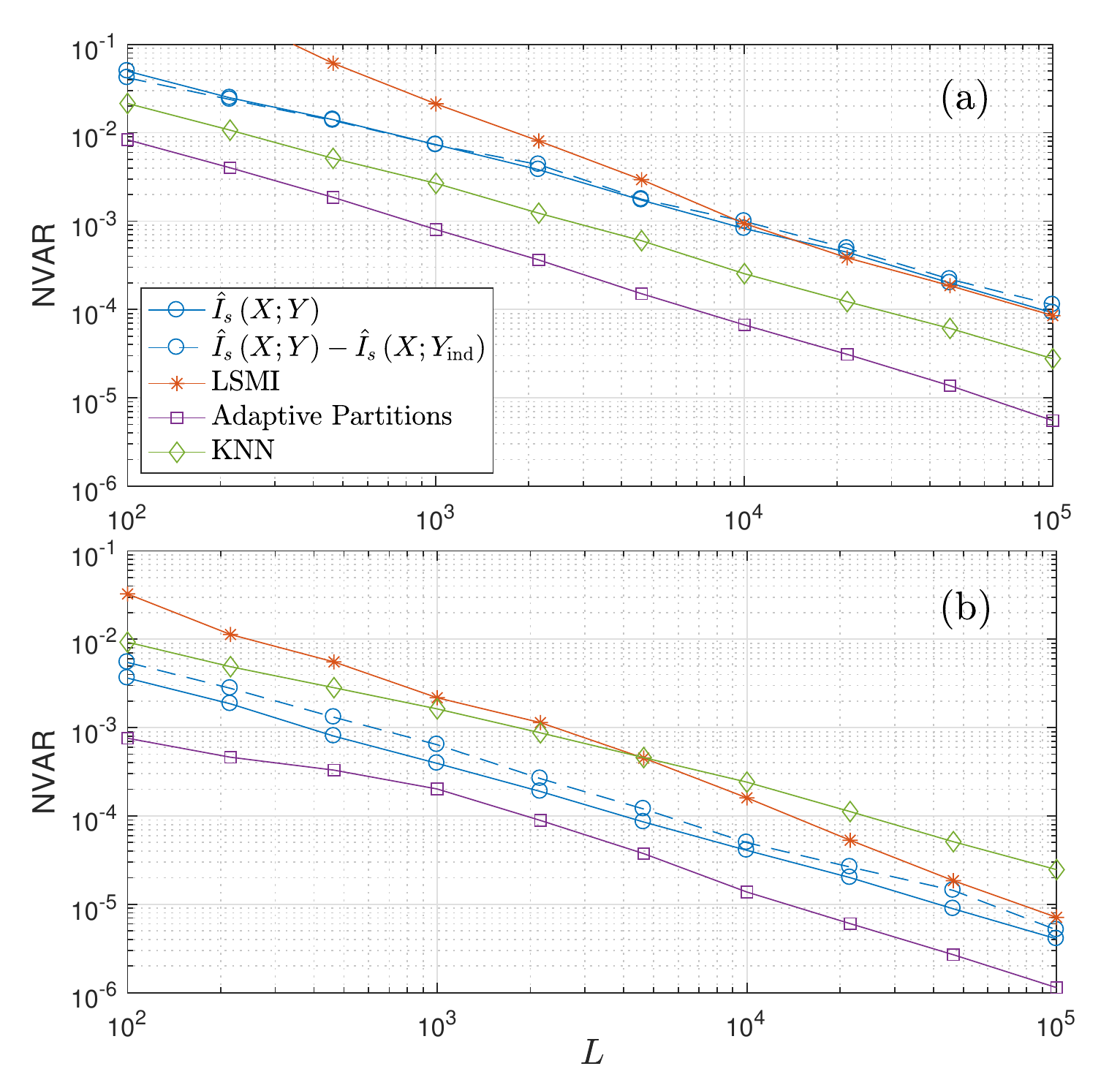}}
\caption{Normalized bias and normalized variance of the estimated SMI as a function of data size $L$ for $\alpha=1/3$, $\sigma^{2}=p/\left(L^{2/5}\right)$ and $N=2\left\lceil 7.5/\sigma\right\rceil +1$.
($a$): SMI=$1$, MI=$0.4$, $p=0.1$. ($b$): SMI=$0.1$, MI=$0.05$, $p=0.25$.\label{fig6}}
\label{fig6}
\end{figure}

Fig. \ref{fig6} depict the bias and variance vs. $L$ of the proposed estimator $\hat{I}_{s}\left(X,Y\right)$ along with its reduced bias version $\hat{I}_{s}\left(X,Y\right)-\hat{I}_{s}\left(X,Y_{\text{ind}}\right)$. For the selection of the perturbation variance, the estimator uses the classical Silverman's rule \cite{silverman-86}, \cite{chen-2015} derived in the context of nonparametric kernel density functionals estimation, which is known to provide consistent results for small dimensional data \cite{principe_2010}. According with this rule, the perturbation variance is let to monotonically decrease with $L$ as $\sigma^{2}=p/\left(L^{2/5}\right)$, being $p$ a free parameter, which is shown to provide a good bias-variance trade-off at all data-size regimes as well as consistency of the estimate for $L\rightarrow\infty$. This relation between data-size and perturbation variance can also be encountered in the context of spectral density estimation after minimizing the MSE with respect to the taper bandwidth \cite{Haley_optimal_bandwidth}, recalling the resemblance between the perturbation based on Gaussian convolutions and the spectral estimation problem. For clarity, the rationale for using this rule also in the context of estimating information is sketched in Appendix D. It can be seen that the reduced bias estimator is especially effective at small values of $L$, approximating the performance of the original method as $L$ increases, at the cost of providing a slightly higher variance. The least squares mutual information estimator (LSMI) \cite{suzuki-2009} is also shown, whose parameters are selected through cross-validation. For completeness, two MI estimators are also tested: one is based on adaptively partitioning the observation space \cite{darbellay-1999}, and the other consists on measuring entropy through the k-nearest neighbor algorithm \cite{krashov-2004} with a single neighbor. Note that, although the true values of SMI and MI differ (both measured through the genie-aided estimator), the comparison between SMI and MI estimators is fair since \textit{normalized} bias and variance are used as performance indicators. 

\begin{figure}[t]
\centering
\includegraphics[scale=0.62]{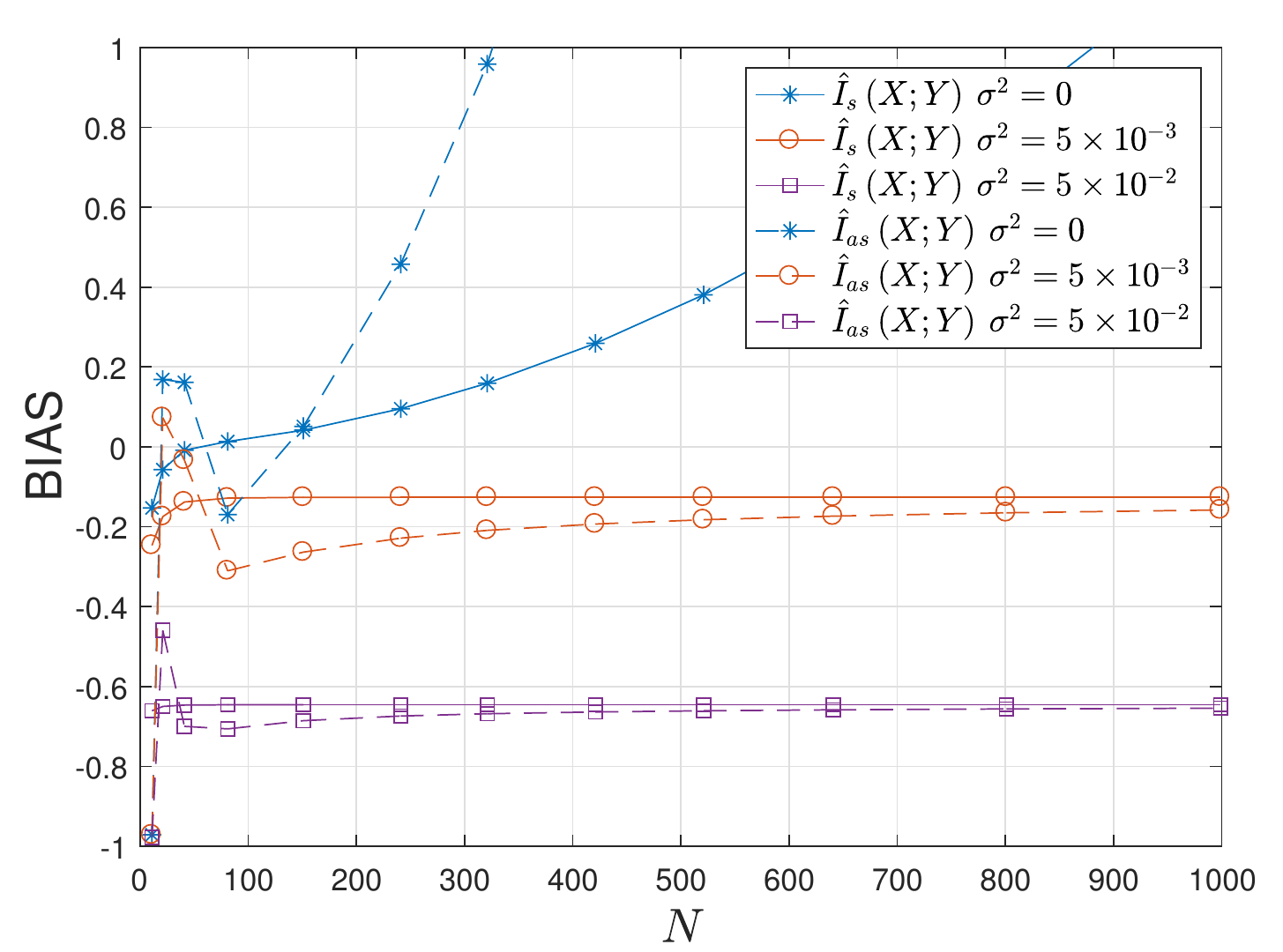}\caption{Bias of the SMI estimator and the approximate SMI estimator as a function of feature space dimension $N$, for $\text{SMI}=1$, $\alpha=1/3$ and $L=10^{5}$.\label{fig8}}
\end{figure}
Finally, the performance of the approximate frequency-domain estimator is shown in Fig. \ref{fig8} in terms of bias. It can be seen that, as the dimension increases, the performance of the approximate estimator converges to that of the original estimator, provided that a nonzero smoothing variance is used, with the advantage of a significantly reduced computational load. Note that the greater is the smoothing variance, the faster is the convergence rate of the frequency-domain estimator to the original performance, at the expense of an increased negative bias.


\section{Conclusion}

In this paper, we have derived estimators of the degree of dependence of a pair of i.i.d. data, which are based solely on second-order statistics computed after mapping the data onto a complex space with higher dimension. The use of second-order statistics is possible as a result of selecting a surrogate of mutual information that is a quadratic measure of dependence. 

In particular, it is shown that the squared-loss mutual information used in the field of machine learning corresponds to a second-order statistics based on the Frobenius norm of a coherence matrix, which is known to be directly linked with the standard CCA tool. The selected squared-loss surrogate has the property of upper-bounding Shannon mutual information. Moreover, it behaves as a local approximation of twice the Shannon mutual information, which means that the developed estimators provide meaningful values at the challenging, small dependence regime. While in the case of discrete data a mapping onto the $(N-1)$-simplex suffices, for analog data the natural feature space is based on steering vectors and its dimension can be selected as a regularization parameter of the problem, trading-off performance (bias) and complexity. The main advantage of avoiding the dual form as in kernel methods is that the estimators become linearly scalable with respect to the data size, and that the free parameters can be selected with physical meaning related to the expected dynamic range and expected smoothing degree of the true densities. In the development of the estimators, some connections with well-known concepts in the literature have emerged, such as the locally optimal detector of correlation for Gaussian data, the linear information coupling problems, the Gebelein maximal correlation, the chi-squared divergence and the spectral analysis.

Finally, some pending issues are left for future work, such as the extension of the estimator to the case of data with memory, as proposed for instance in \cite{malladi-2018}, and a data-dependent dimensionality reduction strategy prior to CCA, for which some preliminary results based on the \textit{minimum description length} principle have recently been provided in \cite{carloslopez-2020}.


\section{Appendices}

\subsection*{Appendix A: Derivation of (\ref{chisq_disc}). }

Defining $P_{X}$ as the probability measure, we have
\begin{equation*}
 D_{\chi^{2}}\left(p_{X}||q_{X}\right)=\int\frac{p_{X}}{q_{X}}dP_{X}-1=\int\left(\frac{p_{X}}{q_{X}}-2\right)dP_{X}+1 =\int\left(\frac{p_{X}^{2}-2p_{X}q_{X}}{p_{X}q_{X}}\right)dP_{X}+\int\left(\frac{q_{X}^{2}}{p_{X}q_{X}}\right)dP_{X}
\end{equation*}
\begin{equation}
=\int\left(\frac{p_{X}^{2}-2p_{X}q_{X}+q_{X}^{2}}{p_{X}q_{X}}\right)dP_{X}=\int\left(\frac{p_{X}-q_{X}}{\sqrt{p_{X}q_{X}}}\right)^{2}dP_{X}=\mathbb{E}_{p_{X}}\left(\frac{p_{X}(x)-q_{X}(x)}{\sqrt{p_{X}(x)q_{X}(x)}}\right)^{2},\label{eq del Ap A}
\end{equation}
as written in (\ref{chisq_disc}).

\subsection*{Appendix B: Proof of Theorem \ref{theo coherence} }

From (\ref{pseudoC}), we have 
\begin{equation}
||\hat{\mathbf{C}}||^{2}=\text{tr}\left(\hat{\mathbf{C}}_{xy}\hat{\mathbf{R}}_{y}^{-1}\hat{\mathbf{C}}_{xy}^{H}\hat{\mathbf{R}}_{x}^{-1}\right),\label{las4}
\end{equation}
where, using (\ref{unas}), $\hat{\mathbf{R}}_{x}=\mathbf{X}\mathbf{X}^{H}/L=\mathbf{F}[\hat{\tilde{\mathbf{p}}}]\mathbf{F}^{H}$, $\hat{\mathbf{R}}_{y}=\mathbf{Y}\mathbf{Y}^{H}/L=\mathbf{G}[\hat{\tilde{\mathbf{q}}}]\mathbf{G}^{H}$ and $\hat{\mathbf{C}}_{xy}=\mathbf{X}\mathbf{P}_{\mathbf{1}}^{\bot}\mathbf{Y}^{H}/L=\mathbf{F}(\hat{\tilde{\mathbf{J}}}-\hat{\tilde{\mathbf{p}}}\hat{\tilde{\mathbf{q}}}^{T})\mathbf{G}^{H}$.
Plugging them on (\ref{las4}), we have 
\begin{equation}
\begin{array}{c}
||\hat{\mathbf{C}}||^{2}=\text{tr}\left(\mathbf{F}(\hat{\tilde{\mathbf{J}}}-\hat{\tilde{\mathbf{p}}}\hat{\tilde{\mathbf{q}}}^{T})\mathbf{G}^{H}\right.(\mathbf{G}[\hat{\tilde{\mathbf{q}}}]\mathbf{G}^{H})^{-1}\left.\mathbf{G}(\hat{\tilde{\mathbf{J}}}-\hat{\tilde{\mathbf{p}}}\hat{\tilde{\mathbf{q}}}^{T})^{T}\mathbf{F}^{H}(\mathbf{F}[\hat{\tilde{\mathbf{p}}}]\mathbf{F}^{H})^{-1}\right)
\end{array}
\end{equation}
Using that $\mathbf{F}$ and $\mathbf{G}$ are invertible, we get
$||\hat{\mathbf{C}}||^{2}=\text{tr}\left(\mathbf{F}(\hat{\tilde{\mathbf{J}}}-\hat{\tilde{\mathbf{p}}}\hat{\tilde{\mathbf{q}}}^{T})[\hat{\tilde{\mathbf{q}}}]^{-1}(\hat{\tilde{\mathbf{J}}}-\hat{\tilde{\mathbf{p}}}\hat{\tilde{\mathbf{q}}}^{T})^{T}[\hat{\tilde{\mathbf{p}}}]^{-1}\mathbf{F}^{-1}\right)$.
Finally, the circularity of trace allows writing 
\begin{equation}
||\hat{\mathbf{C}}||^{2}=\text{tr}\left((\hat{\tilde{\mathbf{J}}}-\hat{\tilde{\mathbf{p}}}\hat{\tilde{\mathbf{q}}}^{T})[\hat{\tilde{\mathbf{q}}}]^{-1}(\hat{\tilde{\mathbf{J}}}-\hat{\tilde{\mathbf{p}}}\hat{\tilde{\mathbf{q}}}^{T})^{T}[\hat{\tilde{\mathbf{p}}}]^{-1}\right)=||\hat{\tilde{\mathbf{C}}}||^{2},\label{partida}
\end{equation}
as we wanted to show. 

\subsection*{Appendix C: Proof of Theorem \ref{theo coherence-2}}

The following properties are used for the proof: $\hat{\tilde{\mathbf{p}}}^{T}\mathbf{1}_{N}=\hat{\tilde{\mathbf{q}}}^{T}\mathbf{1}_{M}=1$, $\hat{\tilde{\mathbf{J}}}\mathbf{1}_{M}=\hat{\tilde{\mathbf{p}}}$, $\mathbf{1}_{N}^{T}\hat{\tilde{\mathbf{J}}}=\hat{\tilde{\mathbf{q}}}^{T}$, $[\hat{\tilde{\mathbf{p}}}]\mathbf{1}_{N}=\hat{\tilde{\mathbf{p}}}$ and $[\hat{\tilde{\mathbf{q}}}]\mathbf{1}_{M}=\hat{\tilde{\mathbf{q}}}$.
Clearly, 
\begin{equation}
\begin{array}{c}
(\hat{\tilde{\mathbf{J}}}-\hat{\tilde{\mathbf{p}}}\hat{\tilde{\mathbf{q}}}^{T})\mathbf{1}_{M}=\mathbf{0}_{N},\\
\mathbf{1}_{N}^{T}(\hat{\tilde{\mathbf{J}}}-\hat{\tilde{\mathbf{p}}}\hat{\tilde{\mathbf{q}}}^{T})=\mathbf{0}_{M}^{T},
\end{array}\label{elcero}
\end{equation}
which means that $\mathbf{1}_{N}$ and $\mathbf{1}_{M}$ are left and right singular vectors of matrix $(\hat{\tilde{\mathbf{J}}}-\hat{\tilde{\mathbf{p}}}\hat{\tilde{\mathbf{q}}}^{T})$, respectively, associated to its null singular value. From (\ref{partida}), then we can write
\begin{equation}
\begin{array}{c}
||\hat{\tilde{\mathbf{C}}}||^{2}=\text{tr}\left((\hat{\tilde{\mathbf{J}}}-\hat{\tilde{\mathbf{p}}}\hat{\tilde{\mathbf{q}}}^{T})([\hat{\tilde{\mathbf{q}}}]^{-1}+\frac{\beta}{1-\beta}\mathbf{1}_{M}\mathbf{1}_{M}^{T})\right.\left.(\hat{\tilde{\mathbf{J}}}-\hat{\tilde{\mathbf{p}}}\hat{\tilde{\mathbf{q}}}^{T})^{T}([\hat{\tilde{\mathbf{p}}}]^{-1}+\frac{\beta}{1-\beta}\mathbf{1}_{N}\mathbf{1}_{N}^{T})\right),
\end{array}
\end{equation}
for any $\beta$. From the Woodbury identity, we can write
\begin{equation*}
\begin{array}{c}
\left([\hat{\tilde{\mathbf{q}}}]^{-1}+\frac{\beta}{1-\beta}\mathbf{1}_{M}\mathbf{1}_{M}^{T}\right)^{-1}=[\hat{\tilde{\mathbf{q}}}]-\frac{[\hat{\tilde{\mathbf{q}}}]\mathbf{1}_{M}\mathbf{1}_{M}^{T}[\hat{\tilde{\mathbf{q}}}]}{\beta+\mathbf{1}_{M}^{T}[\hat{\tilde{\mathbf{q}}}]\mathbf{1}_{M}}=[\hat{\tilde{\mathbf{q}}}]-\beta\hat{\tilde{\mathbf{q}}}\hat{\tilde{\mathbf{q}}}^{T},
\end{array}
\end{equation*}
\begin{equation}
\begin{array}{c}
\left([\hat{\tilde{\mathbf{p}}}]^{-1}+\frac{\beta}{1-\beta}\mathbf{1}_{N}\mathbf{1}_{N}^{T}\right)^{-1}=[\hat{\tilde{\mathbf{p}}}]-\frac{[\hat{\tilde{\mathbf{p}}}]\mathbf{1}_{N}\mathbf{1}_{N}^{T}[\hat{\tilde{\mathbf{p}}}]}{\beta+\mathbf{1}_{N}^{T}[\hat{\tilde{\mathbf{p}}}]\mathbf{1}_{N}}=[\hat{\tilde{\mathbf{p}}}]-\beta\hat{\tilde{\mathbf{p}}}\hat{\tilde{\mathbf{p}}}^{T}.
\end{array}
\end{equation}
Clearly, 
\begin{equation}
\begin{array}{c}
\lim_{\beta\rightarrow1}\left([\hat{\tilde{\mathbf{q}}}]-\beta\hat{\tilde{\mathbf{q}}}\hat{\tilde{\mathbf{q}}}^{T}\right)\mathbf{1}_{M}=\mathbf{0}_{M},\\
\lim_{\beta\rightarrow1}\left([\hat{\tilde{\mathbf{p}}}]-\beta\hat{\tilde{\mathbf{p}}}\hat{\tilde{\mathbf{p}}}^{T}\right)\mathbf{1}_{N}=\mathbf{0}_{N},
\end{array}
\end{equation}
which means that these two matrices, which are sample covariance matrices for $\beta\rightarrow1$, share with matrix $(\hat{\tilde{\mathbf{J}}}-\hat{\tilde{\mathbf{p}}}\hat{\tilde{\mathbf{q}}}^{T})$ (see (\ref{elcero})) the same singular vectors associated to null singular value. Therefore, for the limiting case of $\beta=1$, rank reduction using full-rank matrices $\mathbf{F}\in\mathbb{C}^{N'\times N}$ (with $N'=N-1$) and $\mathbf{G}\in\mathbb{C}^{M'\times M}$ (with $M'=M-1$) is possible, constrained to computing covariance instead of autocorrelation matrices from data, which proves the equality with the SMI. For $N'<N-1$ and/or $M'<M-1$, however, the smallest singular values will be lost, which proves the inequality.

\subsection*{Appendix D: Perturbation variance setting}

For large $L$, the bias and variance of the SMI estimator are given by
\begin{equation}
\text{bias}(\hat{I}_{s})=-O(\sigma^{2})+O(\sigma^{-1}L^{-1})
\end{equation}
\begin{equation}
\text{var}(\hat{I}_{s})=O(\sigma^{-1}L^{-1}).
\end{equation}
The term $O(\sigma^{2})\geq0$ is a result of the data processing inequality and the consistent (with $L$) terms $O(\sigma^{-1}L^{-1})$ decrease with $\sigma$ as a result of (\ref{valor de N}). Both have also been approximately confirmed by simulations for a wide range of scenarios. As the mean squared error is $\text{mse}(\hat{I}_{s})=\text{bias}^{2}(\hat{I}_{s})+\text{var}(\hat{I}_{s})$, the condition $\lim_{L\rightarrow\infty}\sigma^{2}=\lim_{L\rightarrow\infty}\sigma^{-1}L^{-1}=0$ is required to yield $\lim_{L\rightarrow\infty}\text{mse}(\hat{I}_{s})=0$, which moves to choosing $\sigma$ as a monotonically decreasing function of $L$ such that $\sigma^{-1}L^{-1}$ is also monotonically decreasing. Let us adopt a power law $\sigma=O(L^{-\gamma})$, for which the condition $0<\gamma<1$ guarantees the desired convergence given by 
\begin{equation}
\text{mse}(\hat{I}_{s})=O(L^{-\min\left[4\gamma,1-\gamma\right]}).
\end{equation}
Then, the value of $\gamma$ can finally be optimized by the following MiniMax rule:
\begin{equation}
\gamma=\arg\max_{\gamma}\min\left[4\gamma,1-\gamma\right])=\frac{1}{5}
\end{equation}
similarly as the Silverman's rule for kernel smoothing, which implies setting the perturbation variance as $\sigma^{2}=p/\left(L^{2/5}\right)$, being $p$ the new relative free parameter of the estimator. 



\end{document}